\documentstyle[
aps,prb]{revtex}

\begin{document}


\title{Conformations of amphiphilic diblock star copolymers}

\author{Fabio Ganazzoli%
\setcounter{footnote}{1}\thanks{
E-mail: Fabio.Ganazzoli@polimi.it}} 
\address{Dipartimento di Chimica, Politecnico di Milano, via L. Mancinelli 7,
20131 Milano, Italy}
\author{Yuri A. Kuznetsov%
\setcounter{footnote}{2}\thanks{Web page: http://www.ucd.ie/chpca;
E-mail: yuri@ucd.ie }}
\address{Centre for High Performance Computing Applications, 
University College Dublin,Belfield, Dublin 4, Ireland}
\author{
Edward G.~Timoshenko%
\setcounter{footnote}{0}\thanks{Author to 
whom correspondence should be addressed. 
Internet: http://darkstar.ucd.ie; 
E-mail: Edward.Timoshenko@ucd.ie}
}
\address{
Theory and Computation Group,
Department of Chemistry, University College Dublin,
Belfield, Dublin 4, Ireland}


\maketitle

\begin{abstract}
$\left.\right.$
\par\noindent
SUMMARY:

We study conformations assumed by single diblock star copolymers
in a poor solvent by means of the Gaussian variational theory and
Monte Carlo simulation in continuous space. Cases of stars with
internal and external hydrophobic blocks are analysed. 
While in the former case the collapsed state has an obvious micellar shape,
the latter case exhibits two nontrivial conformational structures.
Apart from the equilibrium state of a globular hydrophobic 
core with hydrophilic daisy loops, one also finds here a metastable state 
of outstretched hydrophilic blocks with hydrophobic subglobules 
at their ends. Such a state appears to be rather long--lived during
the kinetics of collapse of a swollen star. The plots of monomer 
densities and other observables computed by both techniques are 
found to be in good agreement with each other.
\end{abstract}

\section{Introduction}\label{sec:intro}

Presently there is a great deal of interest in properties of
amphiphilic copolymers because of 
the variety and extent of conformational transitions 
they show under a temperature or 
pH variation \cite{amphiphile,Israel}.  For instance, as such copolymers 
contain both hydrophobic and hydrophilic (polar) monomers
they can form monomolecular micelles with a hydrophobic core in water, 
as well as inverted micelles in apolar solvents. 
Furthermore, these systems have an interesting analogy to globular 
proteins in the folded state \cite{Dill}. For a discussion of 
similarities between ``coarse--grained'' features of proteins and 
those of amphiphilic polymers we refer to ref. \cite{Halperin}. 

Linear copolymers, in particular linear diblock copolymers, 
would aggregate under certain conditions, giving rise to a 
fascinating range of  self--assembled supramolecular ordering in 
the absence of solvent \cite{Israel,Grosberg}, and therefore 
the main issues of study traditionally have been 
the critical micelle concentration and the solid state morphology. 
On the other hand, the possibility of forming monomolecular 
micelles is greatly increased by using more complicated topologies, 
for instance star or comb polymers \cite{Rouault}.

Random branching of polymers is indeed quite common, but carefully
controlled polymerisation procedure allows one to obtain regular stars,
which present a particularly nice object for theoretical study due to
their symmetry and simplicity.
Star polymers are believed to find a number of future applications for
coating, as additives and possibly in drug delivery systems related
to their low viscosity and other interesting properties.

The aim of the present paper is to investigate the
behaviour of diblock star copolymers comprising both hydrophilic
and hydrophobic monomers within the same arm in dilute solution.  
We shall not consider here 
the case of miktoarm star copolymers comprising chemically homogeneous 
arms made of two or more different monomers \cite{mikto}.

Our study will be carried out by means of both the analytical 
Gaussian self--consistent theory and Monte Carlo simulation
in continuous space. Such combination of techniques would permit us to 
assess the validity  of the analytical theory, which up to now was
independently developed and
used by our two Groups in Milan and Dublin to investigate 
the behaviour of random linear copolymers, though in seemingly
distinct formulations. 
In Appendix of this article we shall demonstrate the equivalence of these two
versions of the theory by relating the free energy expressions in both. 
We would like to emphasise that this general analytical approach 
is the only one, to the best of our knowledge, which can explicitly 
account for the details of intramolecular conformation, thus providing 
more refined characteristics of the polymer than simply the overall 
molecular size or density profile.
The predicted intramolecular conformations may then be compared 
with those from direct computer simulations.

Computer simulations of star polymers (see e.g. refs. \cite{Grest,Freire} 
for recent reviews) have been carried out by a number of techniques, 
ranging from Monte Carlo methods, both on and off lattice, 
to Brownian Dynamics and Molecular Dynamics \cite{Freire}. 
Most of works on block copolymer stars, however, dealt with the molten 
state and the supramolecular organisation under such conditions.  

The only study on diblock star copolymers in dilute solutions
similar to ours, as far as we are aware, was 
carried out by means of a lattice Monte Carlo simulation
in ref. \cite{Nelson}.  
The latter paper appears to contain
a somewhat confusing description of the adopted interactions model
and its status. Judging by the
actually observed star conformations, those correspond to the case in
which both types of units are strongly immiscible with each other and
also both tend to avoid the solvent molecules.
This is not, of course, what one would really call an amphiphilic polymer, 
in which one type of units should ``like'' solvent while the other should not.

Besides, there are quite obvious limitations of the lattice
model as such (see e.g. discussion in ref. \cite{TorusNew}), which 
become even more serious for a star polymer. Indeed, the core
monomer on a lattice has a great difficulty to move at all due to its high
coordination number (equal to the number of arms $f$). The
resulting rejection of all attempted moves
for some kind of motions is called a quasi--nonergodicity, a problem
which is hard to deal with even for linear copolymers. If such a
situation occurs, it practically means that the results of the 
simulation may no longer be trusted.
A number of complicated non--local moves involving the core
monomer have been suggested to alleviate this problem \cite{Molina}.
However, such moves are not permitted in case of Dynamic Monte Carlo needed
for study of, for instance, metastable states.
Moreover, for polymers preserving the integrity of links (i.e. if
chain segments cannot pass through each other) mixing of
local and non--local moves may result in subtle topological effects
leading to a non--uniform sampling of the phase space and eventually
to breaking of the balance condition needed for convergence
of Monte Carlo scheme to the state of equilibrium \cite{TopologyMC}.
For these and other reasons it is important to 
examine the conformations of diblock
star copolymers in continuous space based on the correct model of
hydrophobic--hydrophilic interactions.

The plan of our paper is as follows. 
First, we briefly summarise the analytical approach used.
Then, we present the corresponding results by discussing 
both the molecular  size and the intramolecular conformation. 
Next, we introduce the model for Monte Carlo simulation and 
present the results from such an approach performing
their comparison with those of the analytical theory. 
Finally, we discuss the validity and limitations of the Gaussian approach.

\section{Analytical approach}
\label{sec:analyt}

We consider regular star copolymers formed from $f$ equal arms, 
each consisting of $N/f$ monomers, with a total of $N_{Tot}=N+1$ monomers, 
including also the core monomer.  On each arm, the monomer beads
are sequentially numbered from $1$ to $N/f$ starting from the branch point, 
which is labelled as $0$. 
The beads  are connected by freely--jointed bond 
vectors of unit length oriented outwards.  Furthermore, 
each arm is formed from two blocks of equal lengths, 
each comprising $N/2f$ beads.
One block consists of hydrophobic beads (or, in brief, H-beads, thus forming 
an H-block), while the other block consists of hydrophilic, or polar, 
beads (P-beads and P-block).
A star polymer with the H-block topologically adjacent to the branch point 
(and thus with an outer P-block) will be called an INNER-H star, 
while a star polymer with the opposite connectivity will be called
an OUTER-H star.  
Obviously, the topological location of the H-beads is quite important 
because in water they would experience a mutually attractive interaction, 
which eventually determines the chain conformation, and in particular 
the formation of globular clusters.  
For simplicity, we shall assume that the type of the bead at the branch 
point is the same as of the beads in the inner blocks.

The equilibrium conformation of the polymer is then determined by a 
self--consistent free energy minimisation based on the procedure
employed in refs. \cite{a,b}  for linear random copolymers.
The intramolecular free energy (see Appendix for more details) consists of 
the configurational entropy and
the intramolecular interactions terms. The latter ones include: 
(i) the two--body interactions, which can be attractive, repulsive or even 
vanishing depending on the type of beads involved; (ii) the repulsive 
screened interactions accounting for the effective ``thickness'' of
the beads; (iii) the repulsive three--body interactions.  
The last two contributions are assumed to be independent from the 
type of the interacting beads, unlike the first contribution.  
Following refs. \cite{a,b}, we may write it in 
the Gaussian approximation, as a sum over 
all pairs of beads of the probability density of contacts
multiplied by the interaction constants specifying the sign and strength of 
these interactions.
Thus, the free energy contribution due to the two-body 
interactions in $k_B T$ units may be written as,
\begin{equation}\label{1}
{\cal A}_2 = \sum_{i<j}B_{ij} \left\langle r^2_{ij} \right\rangle^{-3/2},
\end{equation}
where $\left\langle r^2_{ij} \right\rangle$
is the mean--squared distance between beads $i,j$, and $B_{ij}$ are
the interaction constants.  
The positive and negative signs of these constants mean repulsive 
and attractive interactions respectively.  

Furthermore, following refs. \cite{b,c,d,e} we shall assume in addition that,
\begin{equation}\label{2}
B_{ij} = \frac{B_i+B_j}{2},
\end{equation}
where $B_i>0$ indicates a P-bead and $B_i<0$ an H-bead.  
A rationale for the latter equation was recently presented by two of us 
(see Eqs. (8-10) in ref. \cite{f}). This particular parametrisation
of $B_{ij}$ constants results after the effective integration over the solvent
molecules degrees of freedom proceeding from the Hamiltonian with a 
contact monomer--solvent molecules interaction upon the solution 
incompressibility condition.

For convenience, we may ascribe to each bead a value $y_i=\pm 1$, 
specifying its P- or H-character, thereby writing $B_i$ in the form,
\begin{equation}\label{3}
B_i=\sigma\, y_i + B_0.
\end{equation}
Thus, $\sigma$ may be taken as a measure of the copolymer's amphiphilicity,
while $B_0$ characterises the strength of the H-P two--body interactions.  
In the present paper, we take $B_0=0$, so that the latter interactions 
vanish.  
It should be remembered, however, that these beads still experience the 
weaker repulsive interactions (screened and three--body).

Following the general procedure outlined in refs. \cite{a,g,h}
within the framework of Gaussian approximation we can write the 
intramolecular free energy as a unique function of the average scalar products 
among the bond vectors, $\langle\bbox{l}_i \cdot \bbox{l}_j\rangle$,
which are thus the variables subject to optimisation.  
The only difference of a star polymer with a linear chain 
is then, clearly, only in the expression yielding  the mean--squared 
distances between beads (on the same or on different arms) in terms of
this set of the chosen variational variables.  
This expression is given and discussed in detail in ref. \cite{i} and
is not reported here for brevity.

\section{Results from the analytical approach}
\label{sec:analres}

Calculations were carried out for 3-arm star copolymers with 16 
beads per arm and for 4-arm star copolymers with either 
12 or 16 beards per arm.
In this way, we could compare stars with  either the same number of beads 
(i.e., $N_{Tot}=49$ beads) or with the same arm length (i.e., $N/f=16$ beads).
We chose not to impose any symmetry properties upon statistical
averages over the arms. This was done to make sure 
that no asymmetrical conformations were lost (see discussions in ref.
\cite{e}), although the problem becomes more expensive
in terms of computations.  
A description of the numerical free energy minimisation procedure
and the details of all equations expressing various quantities of interest 
for describing the conformation of a star may also be found in ref. \cite{a}.

\subsection{Molecular size}\label{subsec:a}

Upon increasing the degree of amphiphilicity $\sigma$
both INNER-H and OUTER-H stars show clustering of the H-beads into dense 
globules, but with an essential difference.   
Formation of a single globule is easily achieved for an INNER-H star
through a continuous (second order) transition  
due to the initial close vicinity of the mutually attractive H-beads 
thanks to the star connectivity.  
On the contrary, for an OUTER-H star this is much more difficult to achieve,
so that formation of a single globule proceeds through a discontinuous 
(first order) transition.  
Furthermore, for these stars a long lived metastable state 
consisting of $f$ small clusters at the arm ends is also possible.

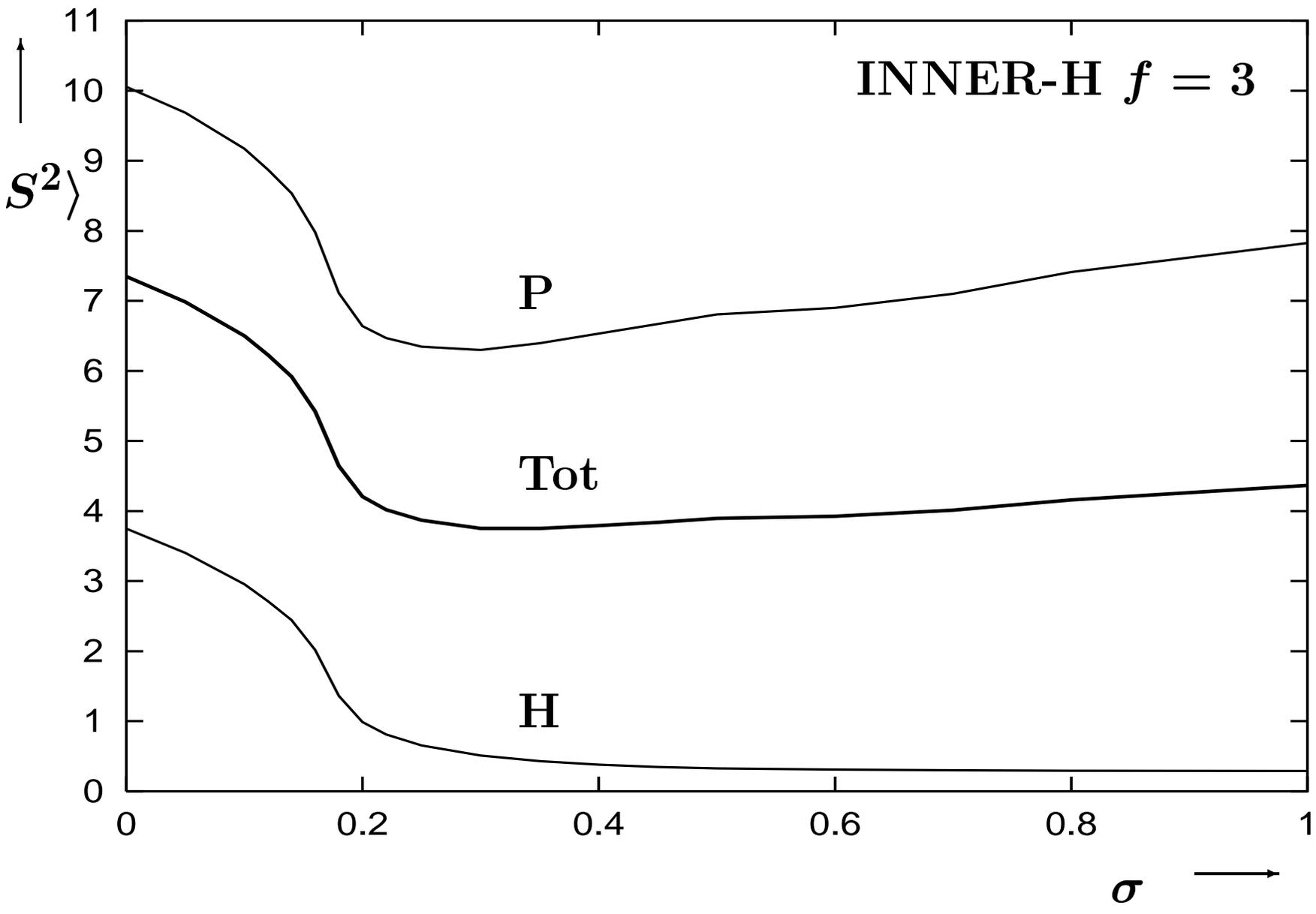
\begin{figure}
\caption{
\label{fig:fa}
The mean--squared radii of gyration of the 3-arm 
INNER-H star copolymer (in units of the bond length $l$) as a 
function of the amphiphilicity parameter $\sigma$.  
The H, P and Tot labels indicate the  mean--squared radius of gyration of 
the H-beads, of the P-beads and of the whole molecule, respectively.
}
\end{figure}

The smooth collapse transition of an INNER-H star can be followed 
through the radii of gyration dependence on $\sigma$, as shown in 
Fig. \ref{fig:fa} for the 3-arm star (results for the 4-arm stars show a
similar pattern and are not shown for brevity).
It is useful to consider the 
mean--squared distance of the H-beads 
from their own centre of mass, $\langle S^2_H \rangle$, which decreases 
monotonously upon increasing $\sigma$ (i.e. increasing attractive 
interactions). At the same time, $\langle S^2_P \rangle$, analogously defined 
for the P-beads, as well as $\langle S^2_{Tot} \rangle$, 
defined as the mean--squared distance of all beads from the 
common centre of mass, show an initial decrease, which then turns
into a steady increase with increasing $\sigma$.  
This behaviour may be easily explained by the increased 
repulsive potential among the P-beads. It should also be noted that  
this trend is in addition enhanced by the pull exerted by 
an increasingly tight core, which brings the arms closer together.  
As a result, the external P-blocks are highly stretched outwards.

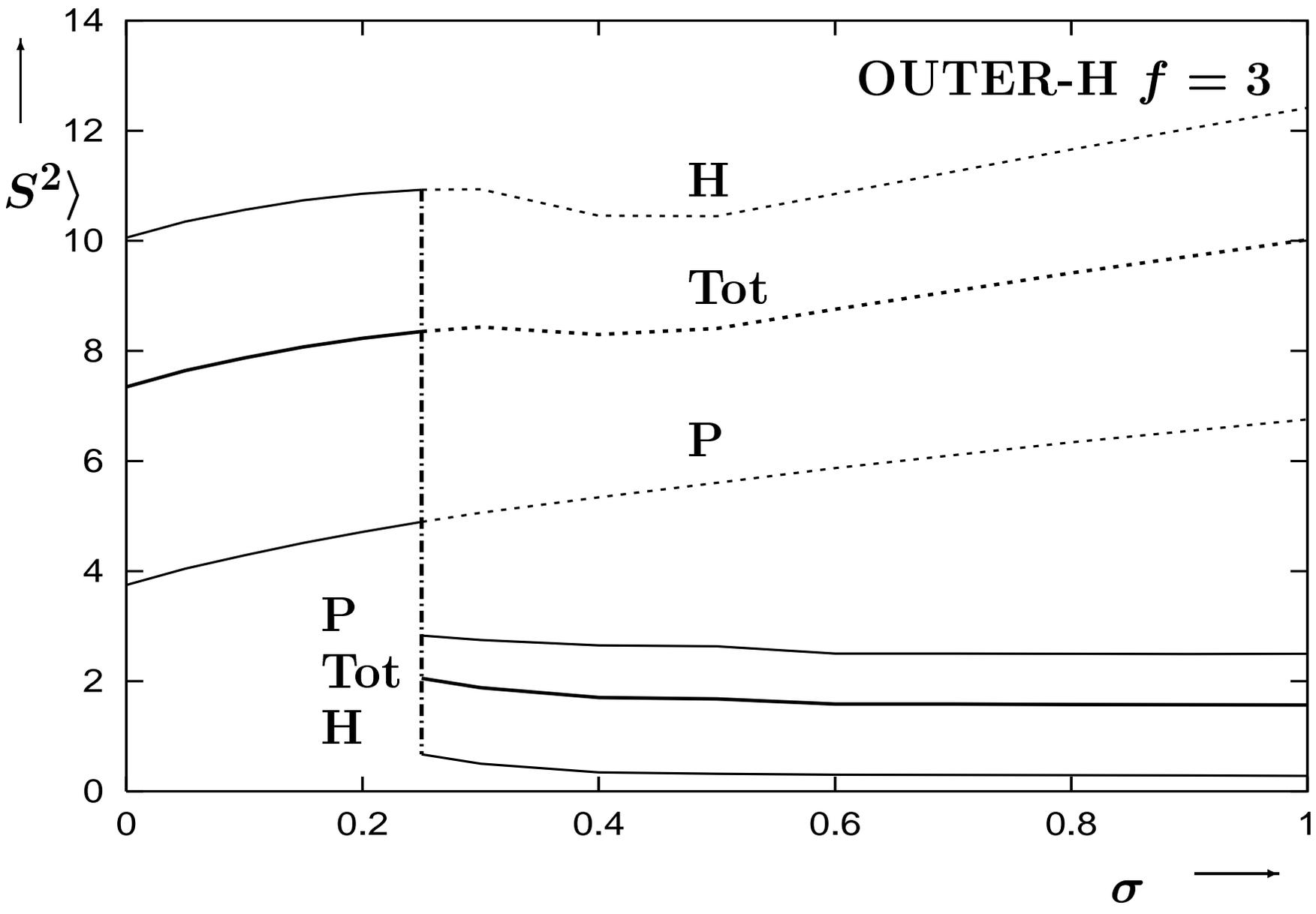
\begin{figure}
\caption{ 
\label{fig:fb}
The same plots as in Fig. \ref{fig:fa}, but for the 3-arm 
OUTER-H star copolymer.  
The dotted lines correspond to the metastable states, and the vertical 
dash--and--dot
line at $\sigma \simeq 0.25$ is the tie line 
connecting the stable states with the same free energy.
}
\end{figure}

An INNER-H star forms a tight globular core easily if the free energy
minimisation uses a random walk conformation as the starting point.
In the case of an OUTER-H star the same procedure, however, only 
leads to a metastable state at large values of $\sigma$. 
Here, the H-beads form $f$ small well separated 
subglobules at the end of each arm.  
Coalescence of these subglobules is prevented by their large 
spatial separation imposed by the inner P-blocks,
which tend to be as far as possible from each other due to 
their mutual repulsion.  
Thus, as shown in Fig. \ref{fig:fb} with dotted curves for the 3-arm star,  
in the metastable state $\langle S^2_P \rangle$  increases 
monotonously with increasing $\sigma$, 
while $\langle S^2_{Tot} \rangle$  
and $\langle S^2_H \rangle$  show a shallow dip when the H-blocks do 
collapse.  Further increase of $\langle S^2_H \rangle$
at larger $\sigma$ is related to the fact that, while the 
individual clusters are actually shrinking, they are being pushed afar 
from the center of mass of the H-beads by the simultaneous 
stretching of the P-blocks.

The stable equilibrium state of an OUTER-H star can only be found 
when using a fully collapsed H-star homopolymer with the same 
total number of beads as the starting point for free energy minimisation.
In this stable state, all H-beads cluster together in a single globule.
This globule also contains the P-bead of the branch point.
The latter feature, however, is not supported by the simulation results
in this paper, showing that in the true stable 
state the core P-bead would actually escape from
the hydrophobic core. Apart from this arguable
behaviour of the branch point, all other P-beads of the 
topologically inner blocks form closed swollen loops with an overall 
daisy--like pattern.  These loops tend to be as far as 
possible from the globular core (as in the stable state of an
INNER-H star), but are seriously constrained by the arm connectivity.  
As a result, in the stable globular state the overall molecular size 
of an OUTER-H star is quite smaller 
than the size of the respective INNER-H star. 

On the other hand, it is interesting to note that the size of 
the dense core of H-beads depends 
only on the total number of beads, but not on the type of star and arms
number, i.e. the same 
$\langle S^2_H \rangle$ values are obtained  
for INNER-H and OUTER-H stars with $3$ or $4$ arms at fixed 
values of $\sigma$ and $N$.
This indicates that the molecular topology becomes irrelevant 
inside dense globules and that their size is dictated only by the 
space filling requirement, in agreement with previous results \cite{StarColl}.

Conversely, for an INNER-H star, using the globular conformation of a 
fully collapsed H-star homopolymer as the starting point of free energy
minimisation results in yet another metastable state.
In such a state, the globular core formed by H-beads has essentially the same 
size as in the stable state, though being less ordered. 
However, the P-beads tend to be much closer to the common centre
of mass, though still protruding outside the globule.  
This conformation was quite unexpected due to the strong repulsive 
interactions experienced by the P-beads.  As a possible rationalisation
for this,  we speculate that such a metastable state encompasses a certain
number of  ``knotted'' arms with  P-blocks being trapped for entropic reasons 
within the H-core.  
We do not report such results in Fig. \ref{fig:fa} since they are not 
supported by the computer simulation results. 

One should realise that the Gaussian variational theory is, in fact,
a mean--field theory which may contain some 
``unimportant'' local minima on the free energy surface. 
These may either be unstable in some directions or separated
by too shallow a barrier from the main minimum, 
which may thus could be easily overcome due to thermal
fluctuations. The latter point is consistent with a relatively poor 
convergence encountered during the numerical free energy minimisation
for this particular metastable state: 
the final mean--squared gradient of the free energy turned out 
to be somewhat larger than expected and all further attempts to improve the 
numerical procedure were plagued by numerical instabilities.

\subsection{Intramolecular conformation}\label{subsec:b}

In order to discuss the intramolecular conformation, 
we shall report some representative results 
obtained for the 3-arm stars with $\sigma$=1.0. 
At large values of $\sigma$ for an INNER-H star the H-beads of the globular 
core experience a strong attractive interaction, which leads to the 
formation of a random globule and further to an 
ordered globule, while the P-blocks stretch outside because of 
their repulsion.  This intramolecular conformation 
is best described via the mean--squared 
distances of the beads belonging to a given arm measured
from the centre of mass of the polymer. This quantity is  
shown in Fig. \ref{fig:fc}a as a function of the position 
along the arm.  We show the plot for only one arm because of 
the statistical symmetry among the arms found in all cases.
The ordering of the hydrophobic core is indicated 
by the zigzag pattern of the H-beads (black circles in Fig. \ref{fig:fc}a), 
which means that the beads lie alternatively on 
two concentric shells centred at the centre of mass.  
Correspondingly, the scalar products among the bond vectors connecting 
the H-beads,  $\langle \bbox{l}_i \cdot\bbox{l}_{i+k}\rangle$ at fixed $i$ 
tend to be equal to $(-1)^k$, as was previously reported for short 
homopolymer chains \cite{g}.  Also, the significantly different 
distances from the centre of mass for the H- and P-beads give rise 
to rather different shapes of the density profiles.  
These density profiles are shown in Fig. \ref{fig:fd} using the
following dimensionless variables,
\begin{equation} \label{rhodef}
\hat{\rho}(r) \equiv 
N_{Tot}^{-1}4\pi r^2 \langle S^2_{Tot}\rangle^{1/2} \rho(r),
\quad 
\hat{r}=r/\langle S^2_{Tot}\rangle^{1/2},
\end{equation}
where $\rho$ stands for any of the three functions $\rho_H$, 
$\rho_P$ or $\rho_{Tot}$.  Note that each 
curve is normalized to the fraction of the corresponding beads within 
the molecule, i.e. $N_a/N_{Tot}$, index $a$ being $H$, $P$, or $Tot$.  

The main features here are the sharp peak for the H-beads
related to their tight clustering in the core, 
and the very broad profile for the P-beads related to their broad 
distance distribution away from the centre of mass.  
As a result, the overall profile of $\rho_{Tot}$ is strongly skewed at 
large values of $r$.
However, because of the molecular connectivity, no more complicated pattern, 
such as say a two peak profile, is observed.

In the case of the 3-arm OUTER-H star, the daisy-like pattern of the 
conformation in the stable state is clearly seen via the mean-squared 
distances of the beads of a given arm from the centre of mass reported in 
Fig. \ref{fig:fc}b (solid line). In particular, the P-beads (white circles)
tend to be as far away from the center of mass as possible, thereby maximising 
their separation, but being constrained by the molecular connectivity 
they return back to the core of the H-beads.  
We should note, however, that according to the present 
results the polar branch point 
(labelled as $i=0$) remains trapped within the hydrophobic core.  
This pecularity though is not supported by computer simulations results
as we have already indicated.

The density profiles for this polymer in Fig. \ref{fig:fe} reflect 
similar features. The density of the H-beads is well localised
with a sharp peak at around $\hat{r}=1/2$ corresponding to a 
tight packing of these beads in the core. On the contrary, the
density of the P-beads is rather delocalised with a shallow
peak around $\hat{r}=1$ corresponding to extended loops of the
P-beads. There is also a slightly non--monotonic behaviour
observed in $\hat{\rho}_P$ at small $\hat{r}$ related to the trapping
of the branch point P-bead within the hydrophobic core.
As for the total density, its profile suggests that overall
the polymer conformation is only dense around the centre of mass
with a fairly long trailing density tail.

Now, if we turn our attention to the conformational structure
of the metastable state of the 3-arm OUTER-H star 
presented in Fig. \ref{fig:fc}b (dashed curve),
we can clearly see an almost linear increase of $\langle R_i^2 \rangle$ with 
arm index for the P-beads (white circles), 
reflecting long and rather stretched P-blocks.
However, as the arm index reaches $1/2$ the
mean--squared distances no longer increase, but remain at the same
value with small zig--zag oscillations corresponding to ordered
small subglobules of the H-beads formed at the ends of the outstretched
arms. As in the case of the stable globular core discussed before, 
the size of these subglobules at a given $\sigma$ depends 
only on the number of H-beads they comprise, but not on the number of arms, 
thus again being dictated by space filling requirement.

Accordingly, the density profiles for this metastable state in 
Fig. \ref{fig:ff}
show that the peak in $\hat{\rho}_P$ occurs at smaller $\hat{r}$ than
in $\hat{\rho}_H$ and it is marginally higher, so that the H-beads are
indeed concentrated at the ends of the P-blocks in
the form of $f$ clusters.
Finally, the total density has a shape similar to
both of the partial densities, though it is about twice higher
and has the width of $\hat{\rho}_H$. 


\section{Continuous space Monte Carlo simulation model}
\label{subsec:cmc}

Many previous simulations of star polymers were performed
on a lattice. However, any lattice model has a number of serious
disadvantages and artefacts. These include 
rotational anisotropy, tendency for condensed phases
to form crystalline structures due to discrete lattice spacing, and
dramatic reduction of the acceptance ratio for moves 
of the core monomer. Thus, we have carried out our simulation
in continuous space, which provides by far a more realistic, though
somewhat more computationally expensive, description.

The model is implemented for a single star polymer consisting of $N_{Tot}$
monomers connected by harmonic springs between any two monomers $i$ and $j$
within arms of the star, which we denote as $i \sim j$. In addition,
all monomers interact with each other via a pair--wise short ranged
spherically symmetric potential, so that the Hamiltonian is
\begin{equation}
H = \frac{k_B T}{2\ell^2} \sum_{i\sim j} r_{ij}^2
  + \frac{1}{2} \sum_{i\not= j} V_{ij} (r_{ij}).
\label{cmc:hamil}
\end{equation}
This bead--and--spring model is much more tractable for a Monte Carlo
simulation than the freely--jointed model adopted in the analytical
part of this work. In any case, since we are interested in rather
generic conformational properties of star copolymers, the differences
between these implementations of the molecular connectivity should
be quite insignificant.

Unlike the Gaussian theory, where for technical reasons one has to introduce a
virial--type expansion representing this pair--wise potential,
here we can use a more realistic two--body interaction potential explicitly,
\begin{equation}
V_{ij}(r) = \left\{
\begin{array}{ll}
+\infty & \quad\mbox{for\ } r < d \\
V_{ij} \left( \left( \frac{d}{r}\right)^{12}
- \left( \frac{d}{r} \right)^{6} \right) & \quad\mbox{for\ } r > d
\end{array}
\right.. \label{cmc:V}
\end{equation}
Thus, monomers are represented by hard spheres of the diameter $d$,
with a short--ranged Lennard--Jones attraction of characteristic
strengths $V_{ij}$. 
The latter coefficients have the expression akin to Eq. (\ref{2}),
\begin{equation}
\label{e2}
V_{ij}=\frac{V_i+V_j}{2},
\end{equation}
where, however, to ensure agreement with Eq. (\ref{3}) we should
take the {\it Ansatz},
\begin{equation}
\label{e3}
V_i=\sigma(1-y_i),
\end{equation}
so that $V_{ij}=0,\sigma,2\sigma$ for any P--P, P--H, and H--H
pair of monomers respectively. 
Thus, although the models employed in this section and in the previous
one are ideologically the same, they do differ somewhat technically.

During simulation 
we change the strength of the degree of amphiphilicity $\sigma$,
which can be viewed as basically the ``inverse temperature'',
rather than changing the temperature $T$ itself.
At $\sigma=0$ there is only a hard core repulsion, which is
represented by a positive three--body coefficient in the virial
expansion of the previous section. 
As a maximal value of the parameter $\sigma$ we chose $\sigma=2.5\, k_B T$,
because in this case the interactions of H--P units
are close to those near the theta--point 
corresponding to vanishing of the second virial coefficient
(this is also consistent with the choice of $B_0=0$ in Eq. (\ref{3})).

The Monte Carlo updates scheme is based on the Metropolis
algorithm with local monomer moves employed by two of us
in ref. \cite{TorusNew}.
The new coordinate of a monomer can be sought as,
$q^{new} = q^{old} + r_{\Delta}$,
where $q$ stands for $x$, $y$ and $z$ spatial projections and
$r_{\Delta}$ is a random number uniformly distributed in the
interval $[-\Delta,\,\Delta]$. Here $\Delta$ is some additional
parameter of the Monte Carlo scheme, which, in a sense,
characterises the timescale involved in the Monte Carlo sweep (MCS),
the latter being defined as $N$ attempted Monte Carlo steps.

In the current model we work in the system of units such that $\ell=1$, 
$k_B T =1$, which introduces respective reduced variables for length and
energy, and in addition we choose the hard core diameter $d = \ell$.

\section{Results from Monte Carlo simulation}
\label{subsec:molres}

Simulations have been carried out for diblock star copolymers
with $f=3,6,9,12$ arms and arm length $N/f=50$ in all cases.
An initial canonical conformation, prepared as a perfect
geometrical star in a 2-d plane, is first
equilibrated by applying a large number of Monte Carlo sweeps
(typically well over $4 N_{Tot}^2$ sweeps)
at $\sigma=0$. This procedure allows one to prepare a 
statistical ensemble (in our case of $Q=50$) statistically
independent homopolymer stars in the swollen coil state.
Even though we formally preserve notations of H- and P-beads,
in this case all beads experience hard--core repulsions.

This initial statistical ensemble is then used for collapsing
each of these stars independently by increasing parameter $\sigma$, so that
the H-beads become increasingly hydrophobic while the P-beads remain
polar, this being done in two possible ways.
First, one could perform a quasistatic process,
in which $\sigma$ is changed in $10$ small steps 
until we reach its maximal possible value $\sigma_{max}=2.5\, k_B T$,
allowing a long equilibration of the star after each parameter change.
In this way we obtain equilibrium conformation for each of these intermediate
values of $\sigma$. Second, we could perform a sudden change 
$\sigma=0\rightarrow \sigma$ and follow the kinetics of the transition
of the initial coil conformation to the final collapsed equilibrium
state at $\sigma_{max}$. Such kinetic procedure is useful for being
able to study formation of possible metastable states, which have
been predicted by the analytical theory above.

As for calculation of statistical averages of various observables
of interest, for the kinetics one is only allowed to average over
values in each of the $Q$ independent members of the statistical
ensemble at the same moment in time $t$, whereas, thanks to the ergodicity
property, for the equilibrium states
one could average over both  the ensemble $Q$ and different
moments in time (i.e. different Monte Carlo sweeps well separated
from each other to improve statistical independence).
The latter allows one to obtain much better statistics and hence smoother
curves in case of equilibrium measurements, while getting the same
quality for the kinetics would seem too time consuming.

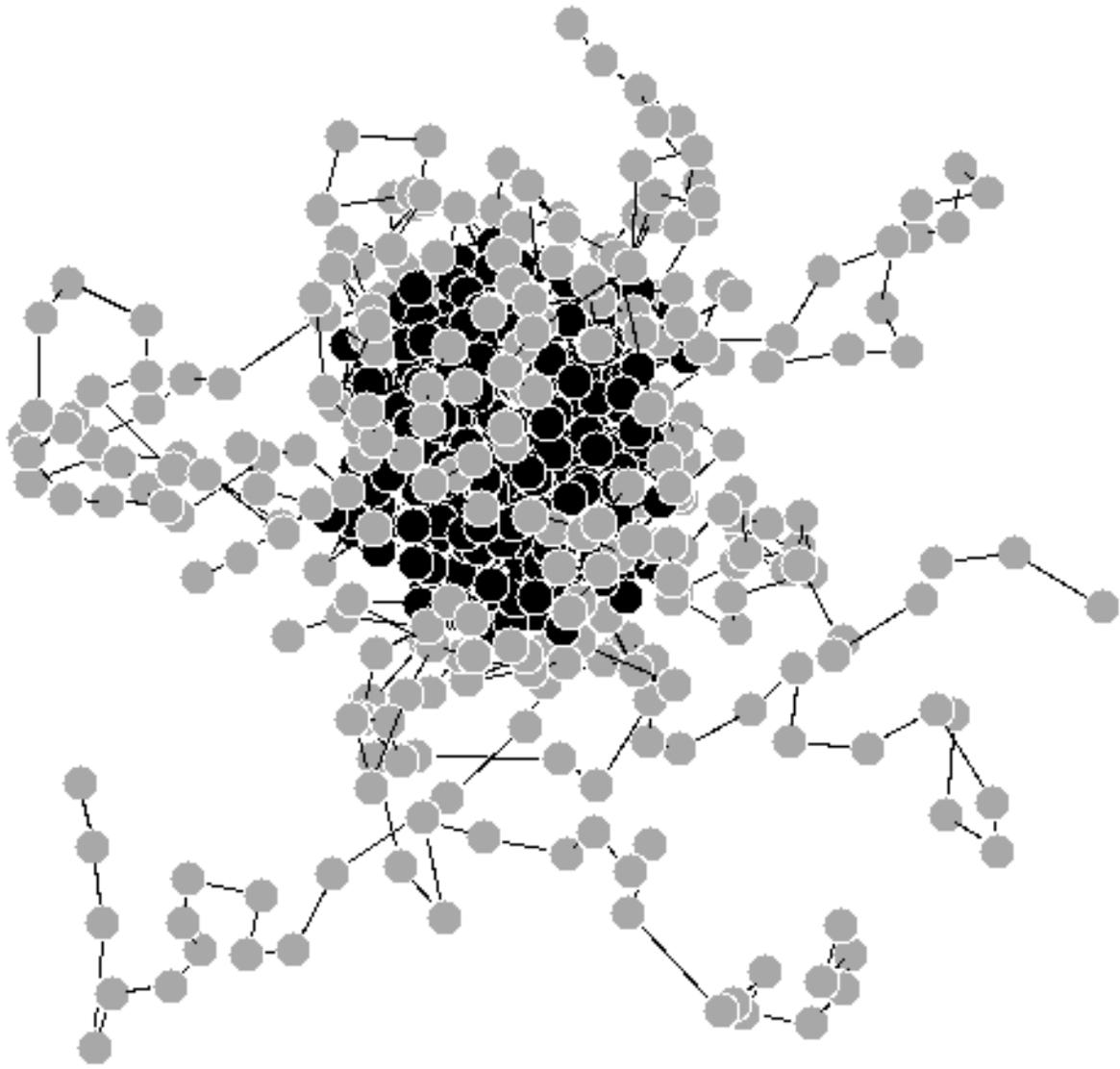
\begin{figure}
\caption{ 
\label{fig:p1}
A snapshot of the conformation of the
INNER-H 12-arm star. Here and in all other figures from the Monte Carlo
simulation  $\sigma=2.5$ and
the colours are assigned as follows: black to H-beads,
grey to P-beads, and clubsuit to the central monomer.
We should also emphasise that the balls size is not up to scale
in these snapshots and has been adjusted simply to look well on paper.
}
\end{figure}

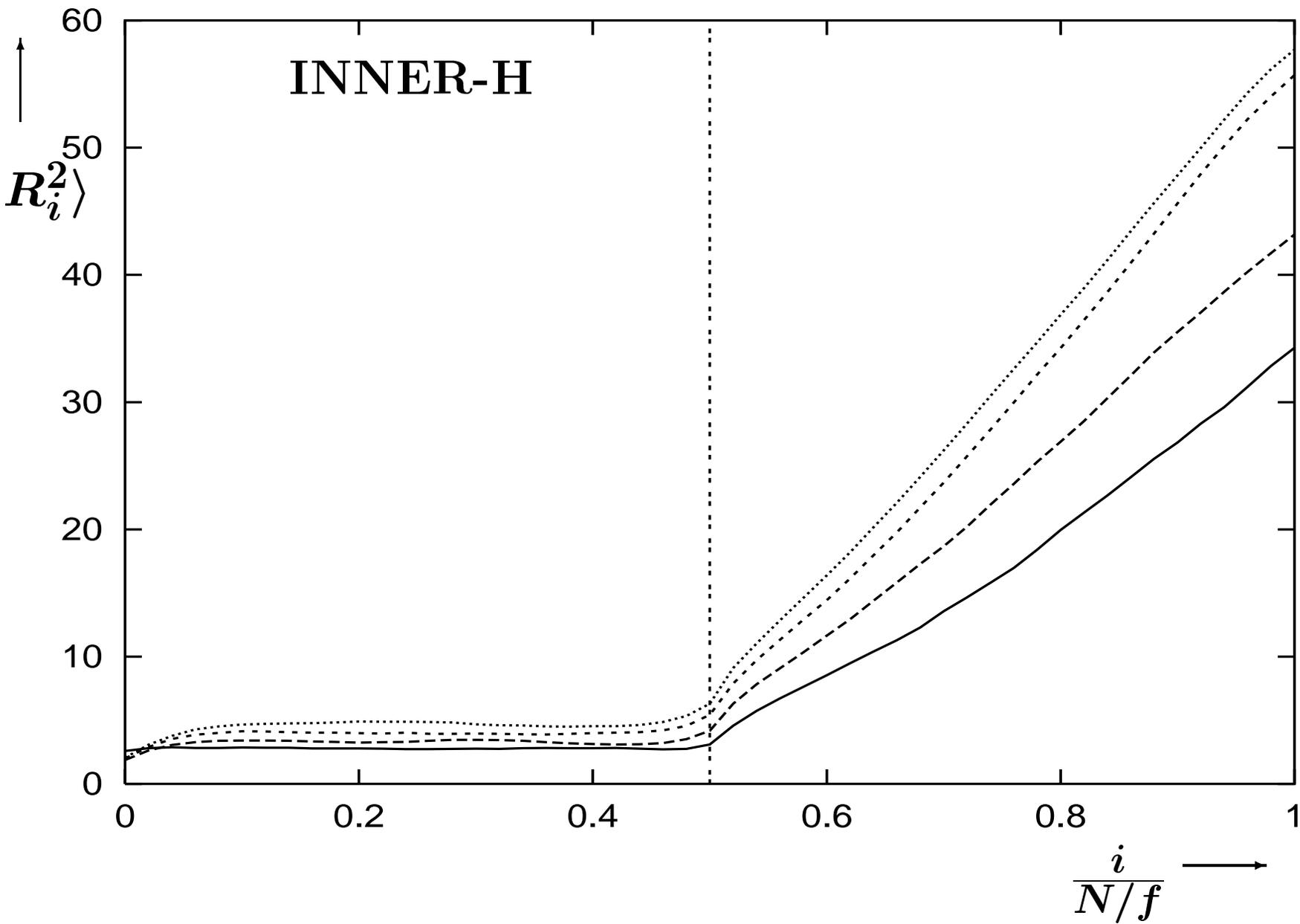
\begin{figure}
\caption{ 
\label{fig:g1}
The mean--squared distances of the beads from the centre of mass, 
$\left\langle R_i^2 \right \rangle$, 
(in units of the bond length $\ell$) vs their position along an arm, $i/(N/f)$,
for the INNER-H stars with $f=3,6,9,12$ (from bottom to top) 
obtained from Monte Carlo simulation.  
Here and below, each star's arm consists of $50$
monomers, and the amphiphilicity parameter is $\sigma=2.5$.
}
\end{figure}

Now, we are going to compare the results from the computer 
simulation with those of the analytical Sec. \ref{sec:analres}.
A typical snapshot of the equilibrium collapsed state of the INNER-H star
is shown in Fig. \ref{fig:p1}. Indeed, there is a nearly spherical
dense core of fully collapsed H-beads (black colour) and rather
outstretched P-blocks (gray colour) hands sticking out of it in
a micellar fashion. Now, if we compare the corresponding plots in Figs. 
\ref{fig:g1} and \ref{fig:fc}a for $\langle R^2_i\rangle$ vs the arm index,
bearing in mind the difference between the theoretical and simulation models
and different sizes of considered stars, the agreement between them seems 
very striking indeed.  We may also observe from Fig. \ref{fig:g1} that the
larger is the star (i.e. greater $f$), the larger is its size and the
higher is the slope of an increasingly linear increase of 
$\langle R^2_i\rangle$ with the arm index for the P-beads.
A zig--zag pattern for the H-beads is absent in Fig. \ref{fig:g1}
because the bead--and--spring model does not lead to any particular
ordering of the globule, contrary to the freely--jointed chain model,
in which bonds have to be arranged in a special manner consistent
with close packing.
Yet, the observed behaviour indicates a rather universal
character of the subject in hand. 

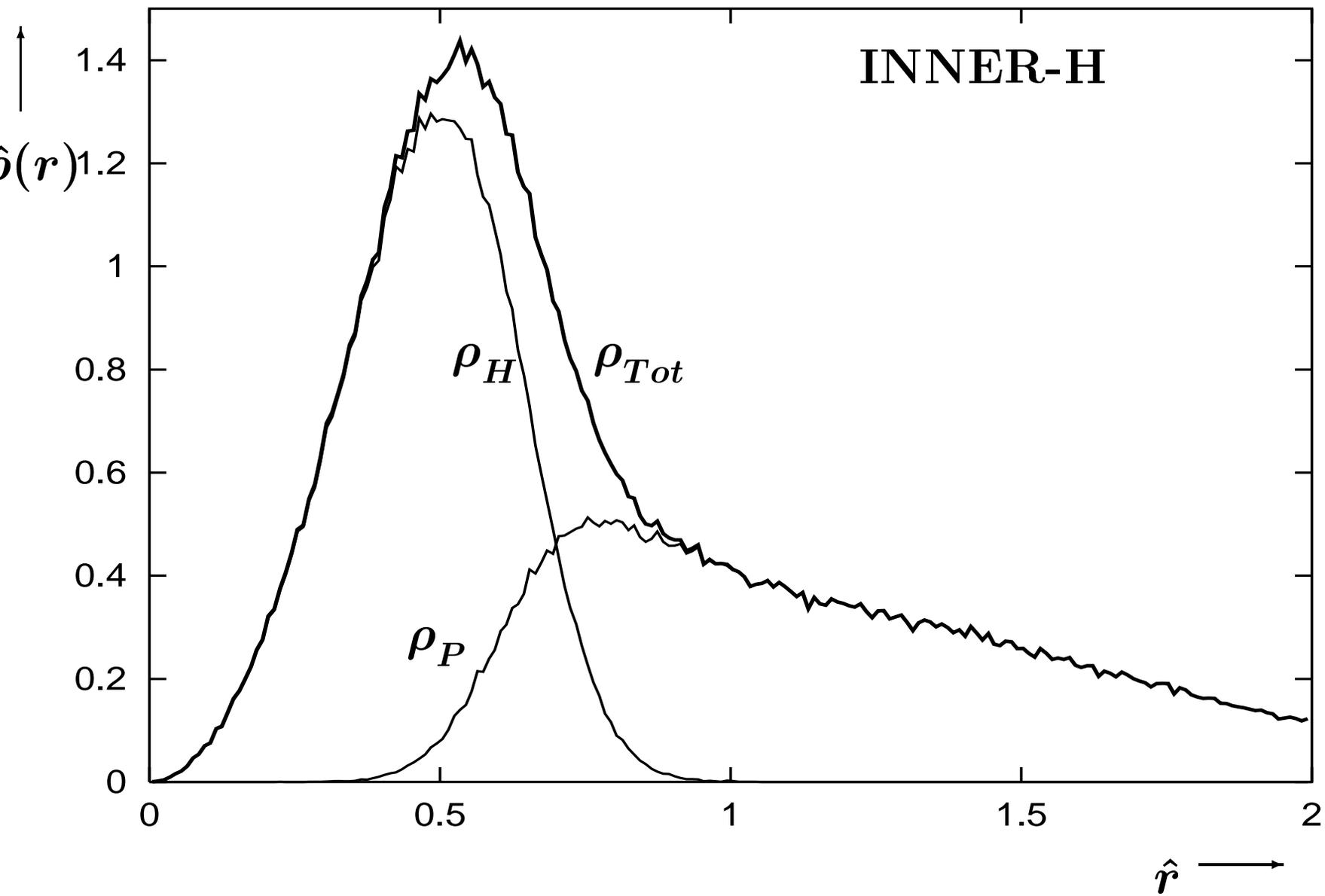
\begin{figure}
\caption{ 
\label{fig:g4}
The dimensionless density profiles $\hat{\rho}$
vs the dimensionless distance $\hat{r}$ for the INNER-H star
with $f=12$ arms obtained from Monte Carlo simulation.  
Here the thin solid curve, the dashed curve and the thick
solid curve correspond to densities of the H-beads, P-beads and all beads
respectively.
}
\end{figure}

Let us now compare partial and total densities in Figs. \ref{fig:g4}
and \ref{fig:fd}. 
Despite the obvious limitations of the Gaussian approximation,
in which densities can only be represented by a sum of Gaussoids, we
still do find a good general agreement between theory and simulations
once we are comparing reduced dimensionless densities $\hat{\rho}$
vs $\hat{r}$.  The overall shapes, as well location of peaks and their
heights, for all three curves appear to be rather consistent.
To avoid any repetition of their discussion already presented in 
Sec. \ref{sec:analres}, we would like to concentrate instead on the
distinctions, thereby clearly establishing possible limitations of the
theory. First of all, a sum of Gaussoids can never produce a rather
flat shape. Thus, the density of the P-beads, $\hat{\rho}_P$, is practically
zero in Fig. \ref{fig:g4} until almost $\hat{r}\simeq 0.4$ (i.e. within
the H-core), while it cannot possibly be so in Fig. \ref{fig:fd} being
a sum of Gaussoids. 
Apart from that feature, for the H-beads $\hat{\rho}_H$ does vanish abruptly, 
while $\hat{\rho}_P$ and $\hat{\rho}_{Tot}$ 
do have long tails, for $\hat{r} > 1$ in both Figs. \ref{fig:g4} and 
\ref{fig:fd}.
To satisfy the normalisation conditions, because of
a somewhat different behaviour of $\hat{\rho}_P$ near the centre of mass,
all densities in the Gaussian theory have to be somewhat deformed
as compared to their precise shape in the computer simulation.
Nevertheless, the density plots seem quite
consistent which each other, and particularly so for the total density.

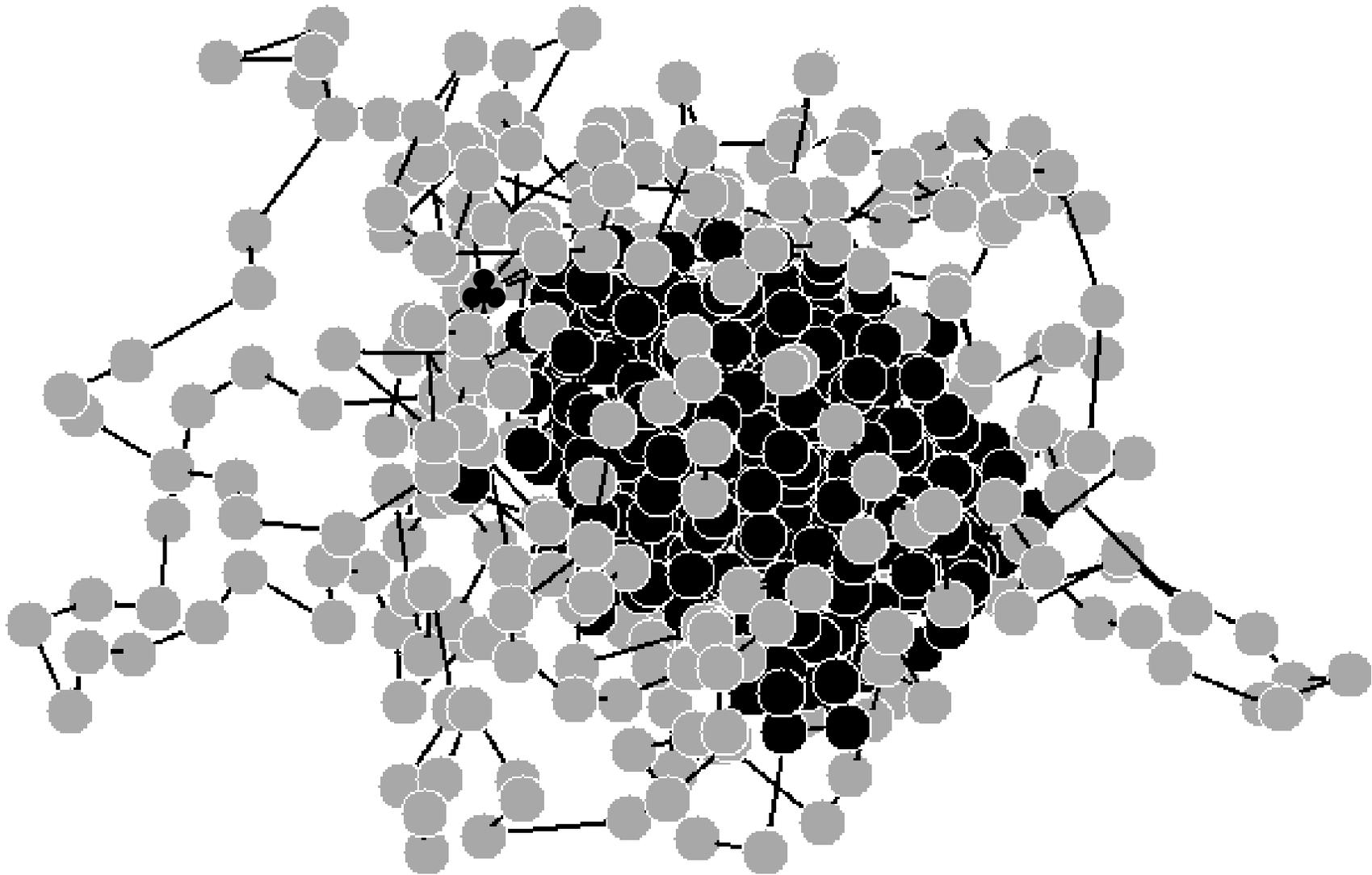
\begin{figure}
\caption{ 
\label{fig:p2}
A snapshot of the conformation  of the
OUTER-H 12-arm star in the stable state at $\sigma=2.5$.
The central monomer is clearly visible here on the outside.
}
\end{figure}

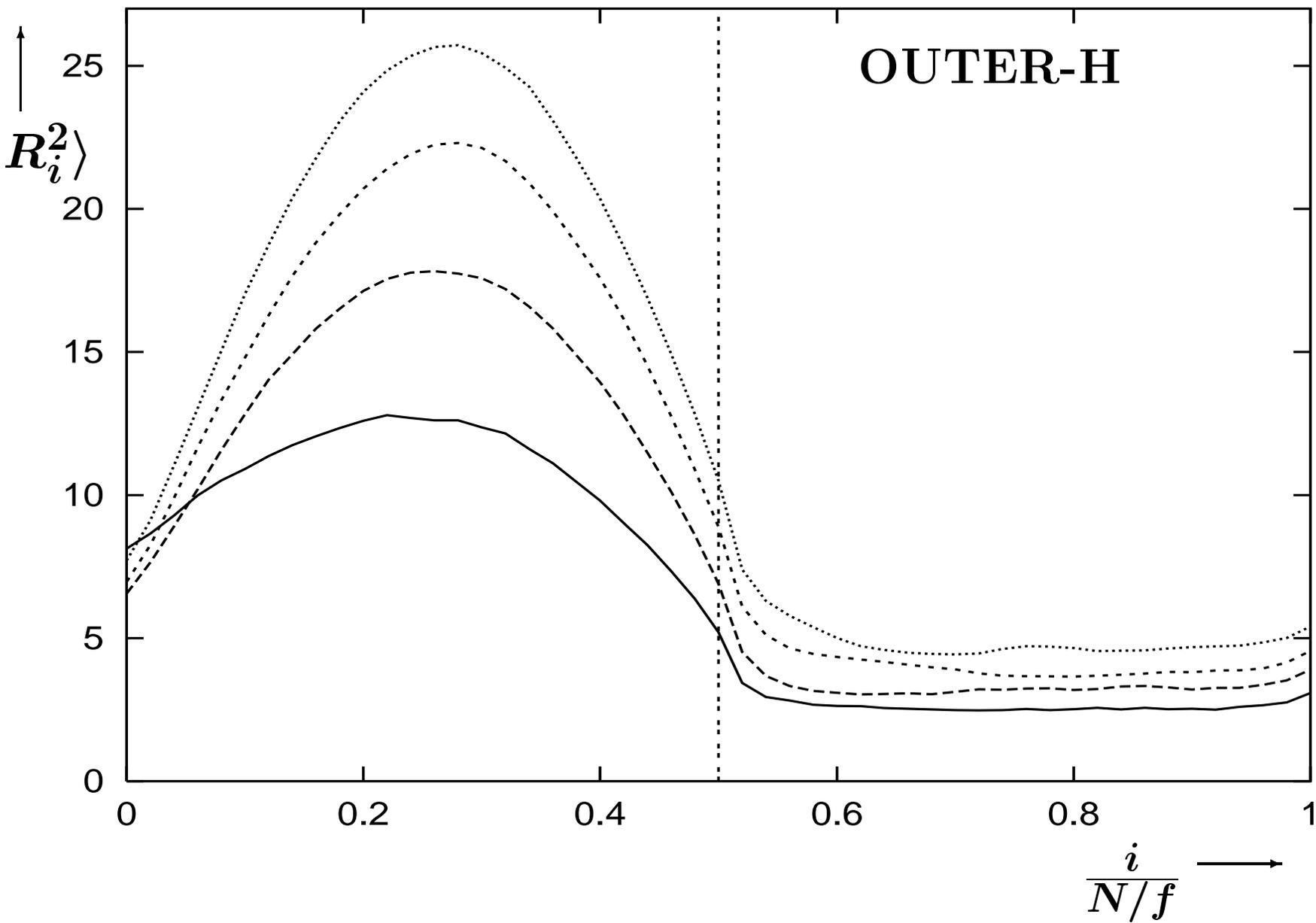
\begin{figure}
\caption{ 
\label{fig:g2}
The plots of $\left\langle R_i^2 \right \rangle$ vs $i/(N/f)$ drawn as 
in Fig. \ref{fig:g1}, but for OUTER-H stars in the stable state,
obtained from Monte Carlo simulation.  
}
\end{figure}

Now then, we turn our attention to the OUTER-H stars.
In Fig. \ref{fig:p2} we exhibit a snapshot of the 12-arm star
in the stable equilibrium state obtained by a quasistatic change
of $\sigma$ with attendant long equilibration after every step. 
The structure does indeed have a nearly spherical collapsed
H-core (black beads), with the P-blocks (gray beads) looping
out in a daisy--like manner. Clearly, the branch point
P-bead (black clubsuit) has been able to escape outside the core
in all of the many analysed members of the ensemble, and it is
the centre to which all other P-blocks are attached by harmonic
springs. Overall structure is much more compact than in Fig. \ref{fig:p1}
and this is also reflected in the graphs in Fig. \ref{fig:g2}
compared to those of Fig. \ref{fig:fc}b. In order to avoid
over--emphasising the good general agreement again, let us instead remark
that the branch P-bead is trapped around the centre of mass in
the analytical result of Fig. \ref{fig:fc}b, producing too small
values of $\langle R^2_i\rangle$ for neighbouring to it P-beads
as opposed to the simulation results in Fig. \ref{fig:p2}.
In the latter, all P-beads are clearly farther away from the
centre of mass than any of the H-beads, and this behaviour seems
to make more sense physically. Thanks to the availability
of data for different values of $f$, one can also see increasingly
better ability of looping of the P-blocks in stars 
with larger number of arms,
so that the $\langle R^2_i\rangle$ plot becomes more and more
symmetrically bell--shaped on the left.

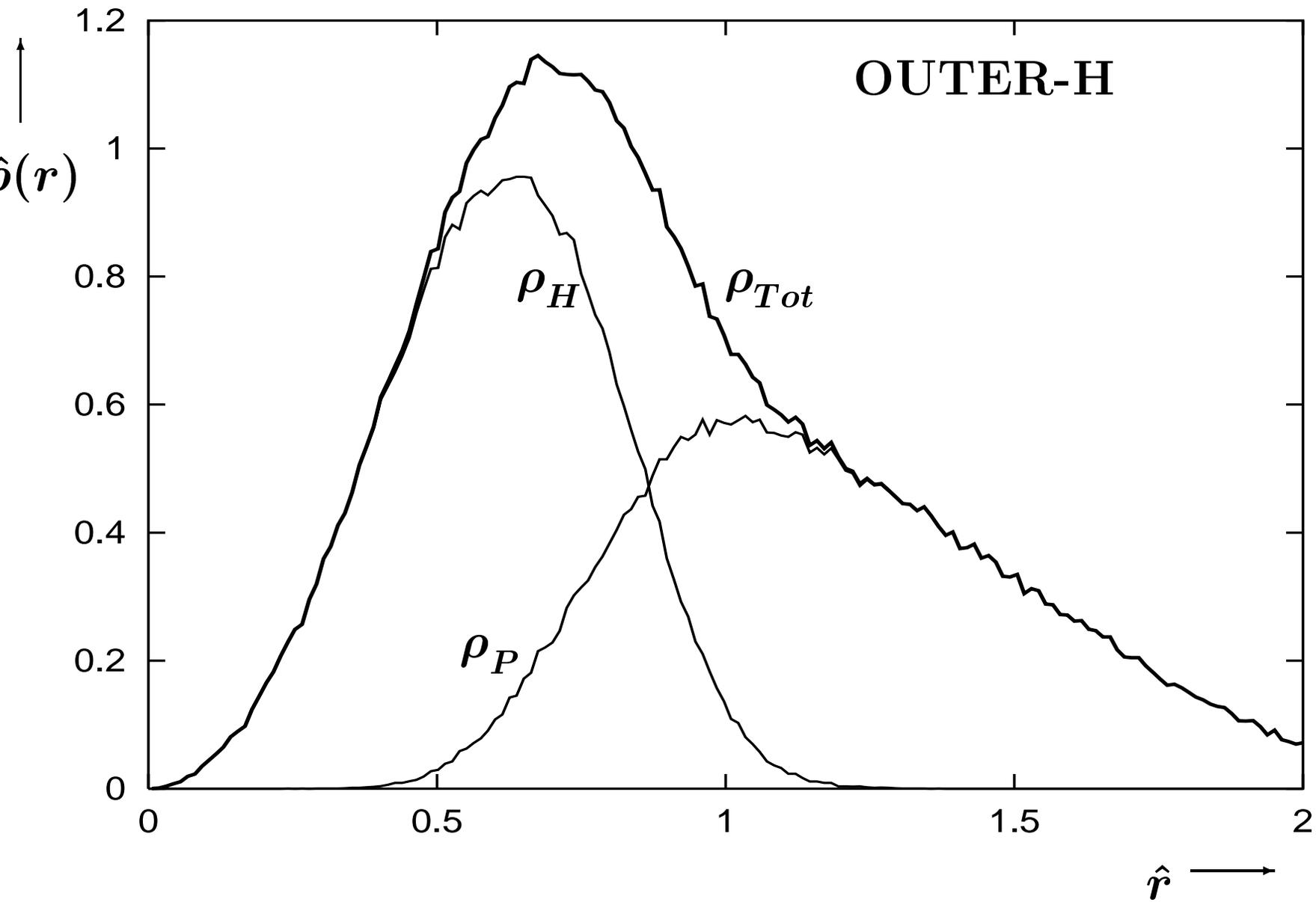
\begin{figure}
\caption{ 
\label{fig:g5}
The profiles of $\hat{\rho}$ vs $\hat{r}$ as in Fig. \ref{fig:g4},
but for OUTER-H star in the stable state, 
obtained from Monte Carlo simulation.  
The same curve notations apply here also.
}
\end{figure}

Density profiles in Figs. \ref{fig:g5} and \ref{fig:fe} again
have consistent overall shapes, peak locations and heights,
except that $\hat{\rho}_P$ is vanishing for $\hat{r}<0.5$ 
in the simulation results and it cannot possibly be so
in the Gaussian theory for the reason already outlined.

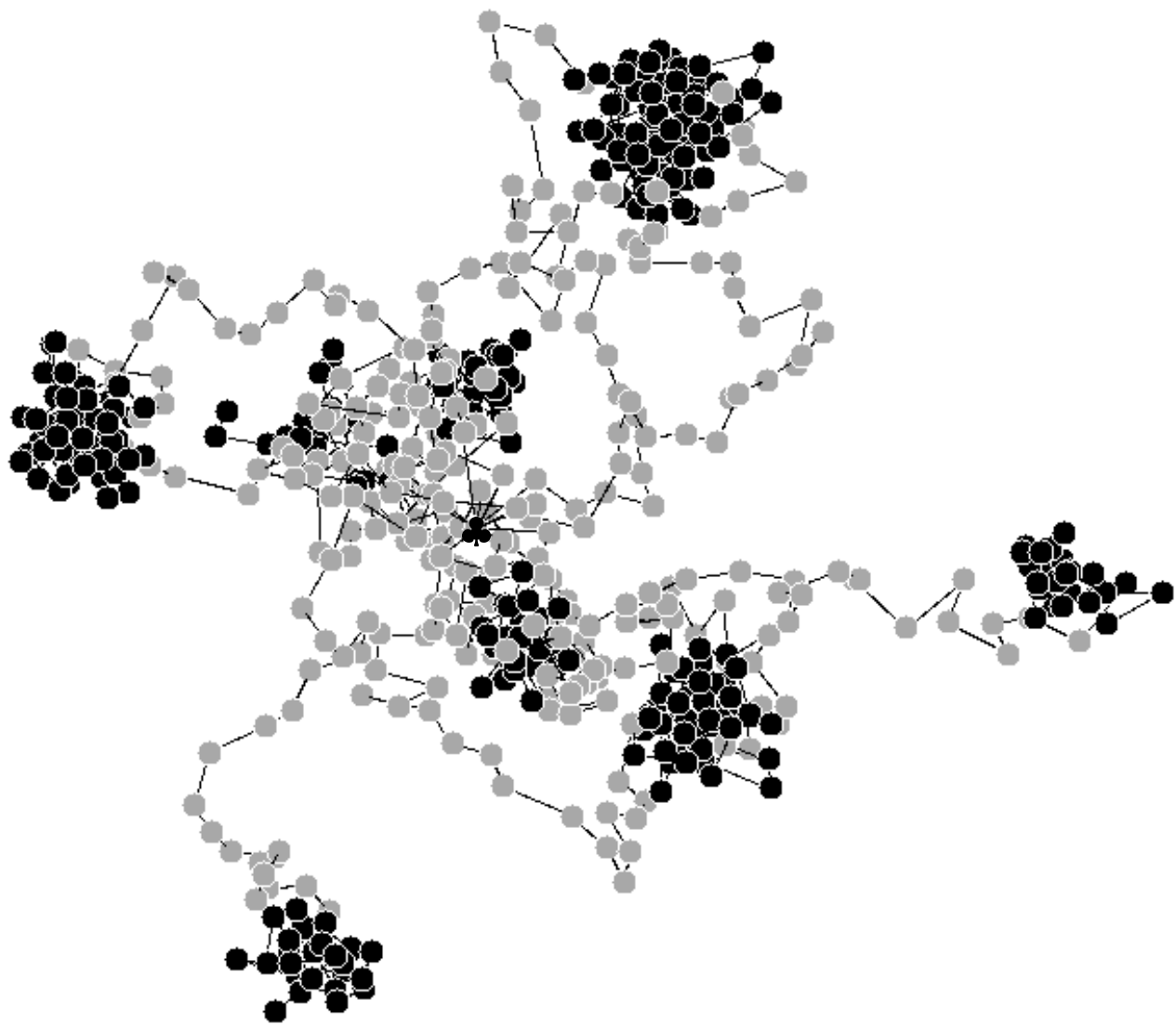
\begin{figure}
\caption{ 
\label{fig:p3}
A snapshot of the conformation  of the
OUTER-H 12-arm star in the metastable state at $\sigma=2.5$.
Note than not all of $f=12$ arms may be well distinguishable here in
a 2-d projection of the 3-d conformation of the polymer.
}
\end{figure}

Now, let us examine the state obtained as a result of the kinetic
process of star collapse after a sudden quench of $\sigma = 0 \rightarrow
\sigma_{max}$. The corresponding snapshot is presented in Fig. \ref{fig:p3}
and it does look like a star consisting of $f$ P-arms with 
small H-subglobules at their ends, as predicted by the analytical theory.
This state was found to be a rather long--lived metastable one
for quenches to $\sigma_{max}\leq 2.5$ --- the associated nucleation time
needed to overcome the barrier for transformation to the equilibrium
state of Fig. \ref{fig:p2} was clearly much longer than the Monte
Carlo times practically accessible to our computational resources (which
consisted of a 16-node Beowulf PII-400 MHz cluster).
However, we have also found that for much deeper quenches the nucleation
regime would transform into the spinodal decomposition one,
in which the state of Fig. \ref{fig:p3} becomes unstable, so that
the kinetics quickly passes through this intermediary to achieve 
the final equilibrium structure.

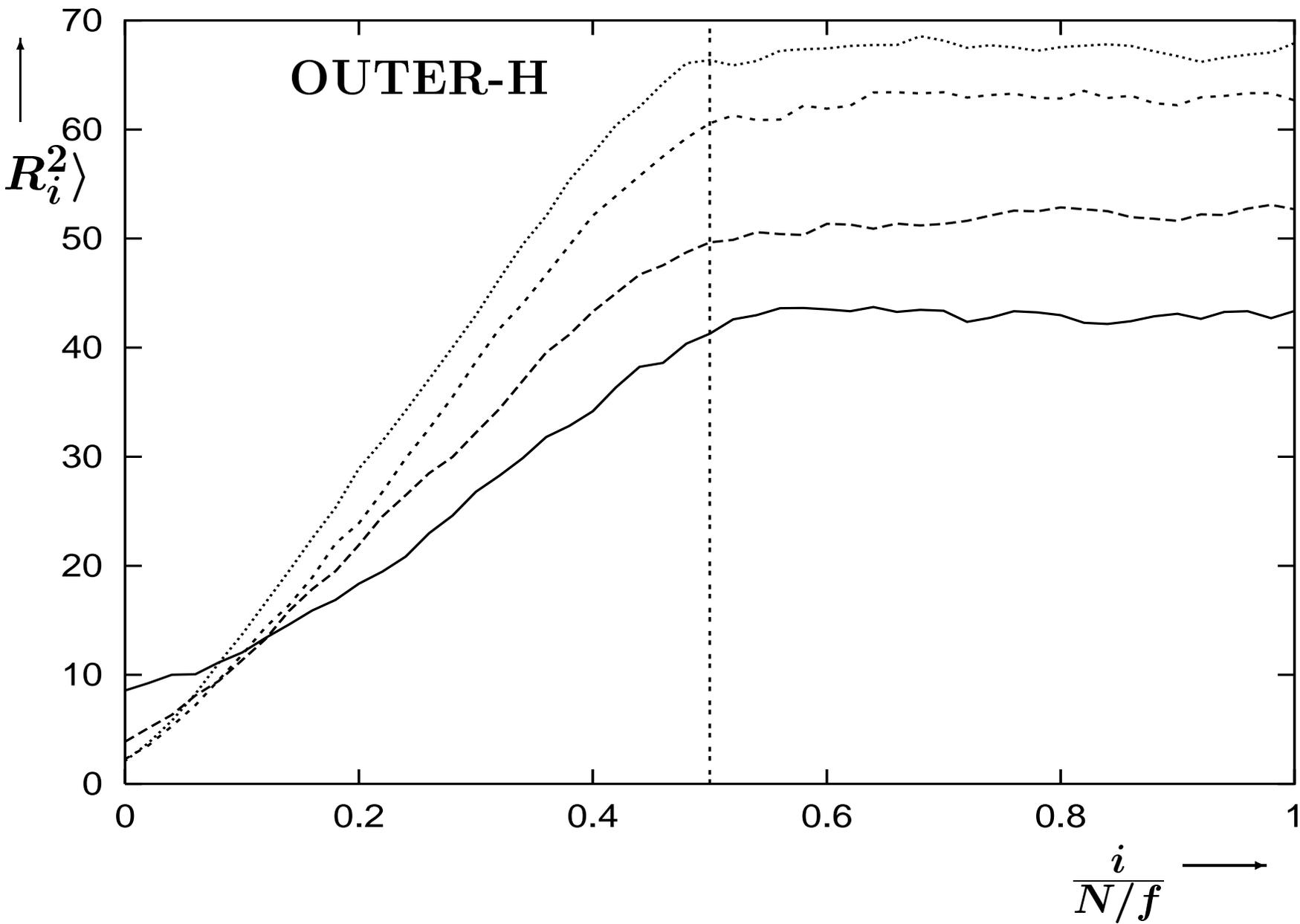
\begin{figure}
\caption{ 
\label{fig:g3}
The plots of $\left\langle R_i^2 \right \rangle$ vs $i/(N/f)$ drawn as 
in Fig. \ref{fig:g1}, but for OUTER-H stars in the metastable state,
obtained from Monte Carlo simulation.  
}
\end{figure}

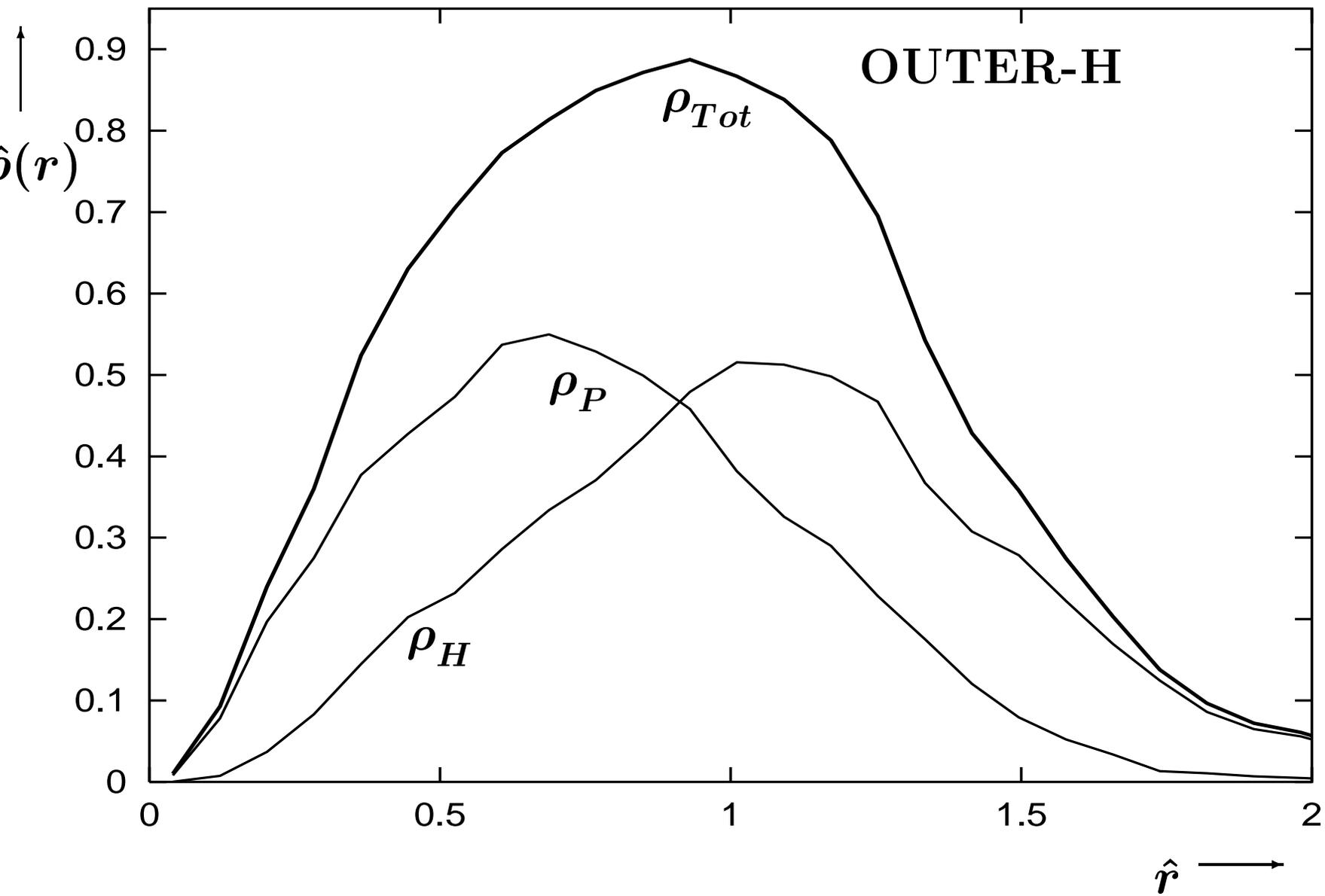
\begin{figure}
\caption{ 
\label{fig:g6}
The profiles of $\hat{\rho}$ vs $\hat{r}$ as in Fig. \ref{fig:g4},
but for OUTER-H star in the metastable state, 
obtained from Monte Carlo simulation.  
The same curve notations also apply.
}
\end{figure}

Looking at the observables such as the mean squared distances
and densities as measured from the centre of mass, there is a good
agreement of Figs. \ref{fig:g3} and \ref{fig:fc}b, as well as 
Figs. \ref{fig:g6} and \ref{fig:ff}, even though in the one case 
a kinetic procedure was used, while a local free energy minimum
was used in the other. We note also that the statistical quality
of Figs. \ref{fig:g3} and \ref{fig:g6} is poorer than of similar
equilibrium plots presented before for obvious reasons, yet
it is quite satisfactory to resolve the shapes well.
The kinetically obtained densities are somewhat broader,
which possibly reflects a better ability of P- and H- beads to move and
mix within their preferred regions of space.

These observations allow us to firmly conclude that the current 
metastable state
of OUTER-H stars is indeed a robust metastable state under the considered
values of the degree of amphiphilicity $\sigma$, while it becomes
an unstable kinetic intermediary for deeper quenches, consistent with
the general theory of phase transitions and many of concrete
examples of similar behaviour previously seen by us for conformational
transitions of other polymeric systems \cite{TorusNew,e}.

Finally, it would be interesting to study the effect of
changing the composition ratio of H:P monomers within the arms
at a given arms number, for instance $f=6$.
Thus, we have studied the following ratios $N_{H}/N_{Tot}=0.2,0.4,0.6,0.8$
as well as the homopolymers $N_{H}/N_{Tot}=0,1$
in addition to the cases of symmetric diblocks $N_{H}/N_{Tot}=0.5$ 
studied for both INNER-H and OUTER-H topologies. 
First of all, we find that the above general conclusions about
the existence of a metastable state for OUTER-H stars as well
as the structures of conformations remain valid in a rather broad
range of composition ratios. 

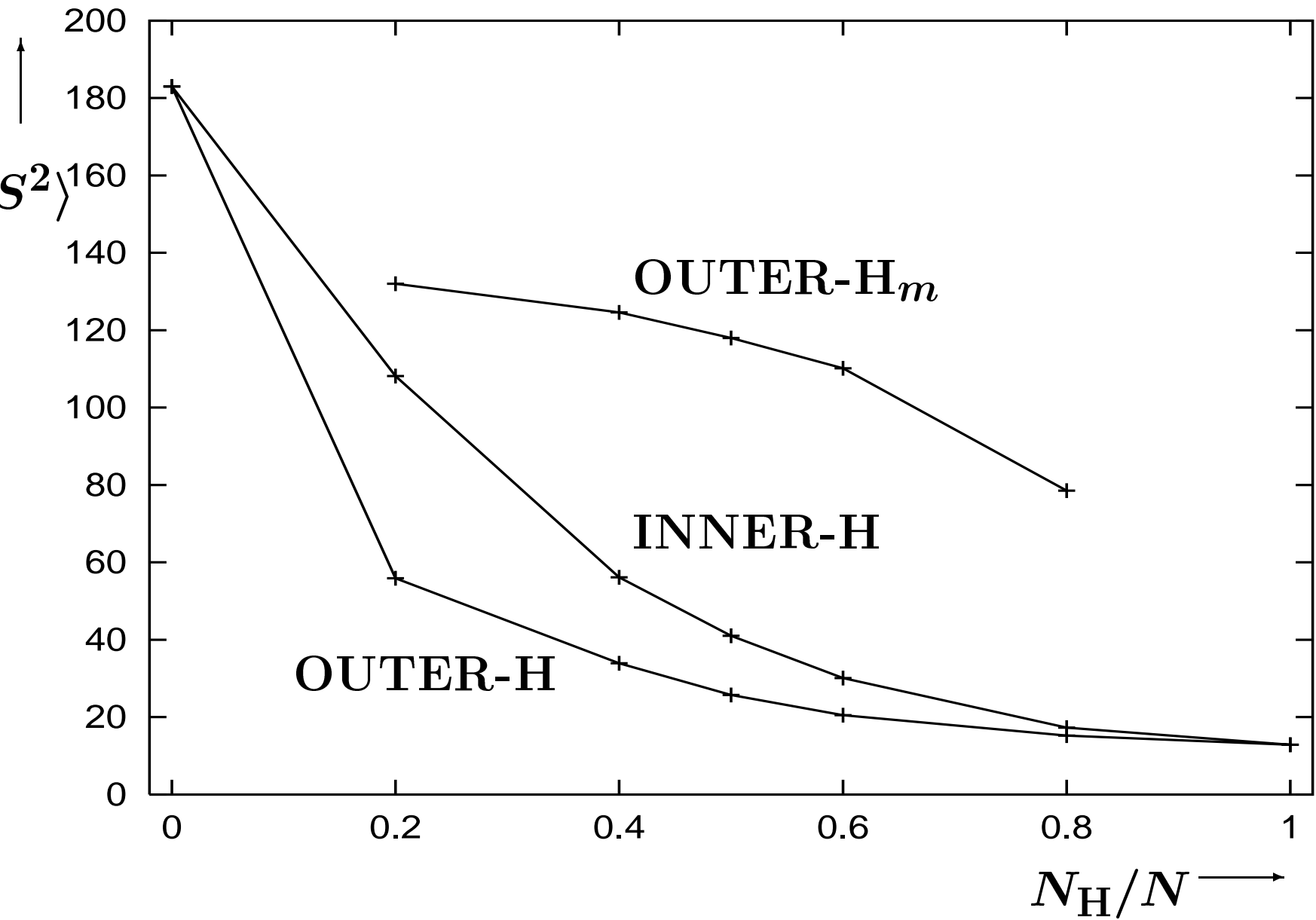
\begin{figure}
\caption{
\label{fig:as1}
Plot of the mean squared radius of gyration, $\langle S^2\rangle$,
vs the composition ratio of hydrophobic units, $N_{{\rm H}}/N$
obtained from Monte Carlo simulation.
Here and below, each star consists of $f = 6$ arms.
Note that the leftmost and rightmost points correspond to homopolymer
stars consisting of hydrophilic and hydrophobic monomers respectively.
The curve denoted by OUTER-H${}_m$ corresponds to the metastable
state.
}
\end{figure}

In Fig. \ref{fig:as1} we present the dependence of the
mean--squared radius of gyration on the H:P ratio 
for the three respective structures in poor solvent.
Obviously, the size of all three structures decreases monotonically
with increasing number of the H-beads. Our conclusion that the
OUTER-H globules are more compact than the INNER-H ones remains
true here throughout, being most pronounced around small
values of $N_{H}/N_{Tot}$, in this case near $0.2$. Interestingly,
the large-size metastable state for OUTER-H stars exists for all 
concentrations except very near the two limiting homopolymer cases.
Moreover, no other metastable states have been seen in kinetics.

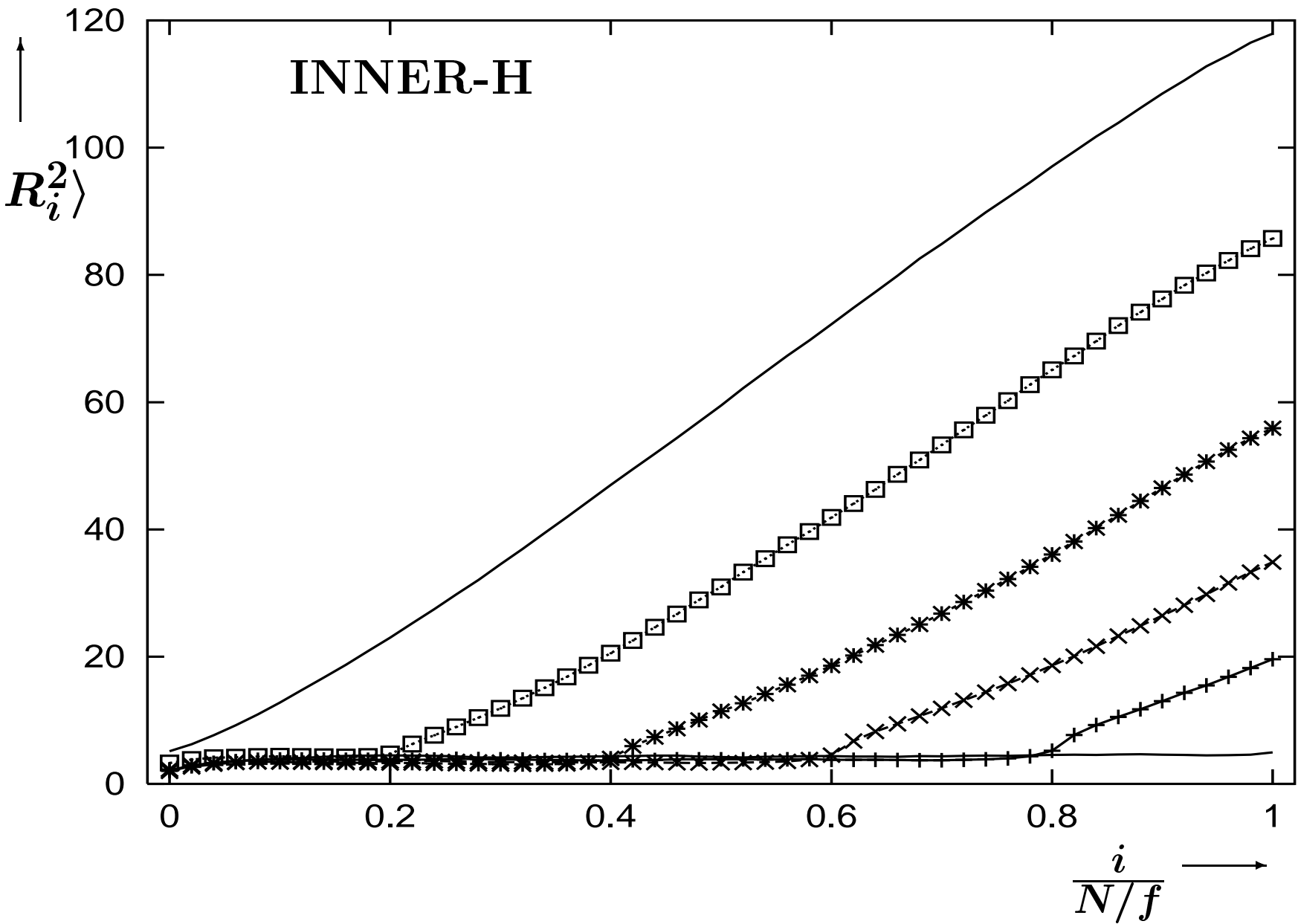
\begin{figure}
\caption{
\label{fig:as2}
The mean--squared distances of the beads from the centre of mass,
$\left\langle R_i^2 \right \rangle$, (in units of the bond length $\ell$)
vs their position along an arm, $i/(N/f)$, for INNER-H stars.
Solid lines correspond to homopolymer stars consisting of hydrophilic
(upper curve) and hydrophobic (lower curve) monomers.
Here and below other lines correspond to stars with
the composition ratios $N_{{\rm H}}/N = 0.2$, $0.4$, $0.6$ and $0.8$
(from top to bottom).
}
\end{figure}

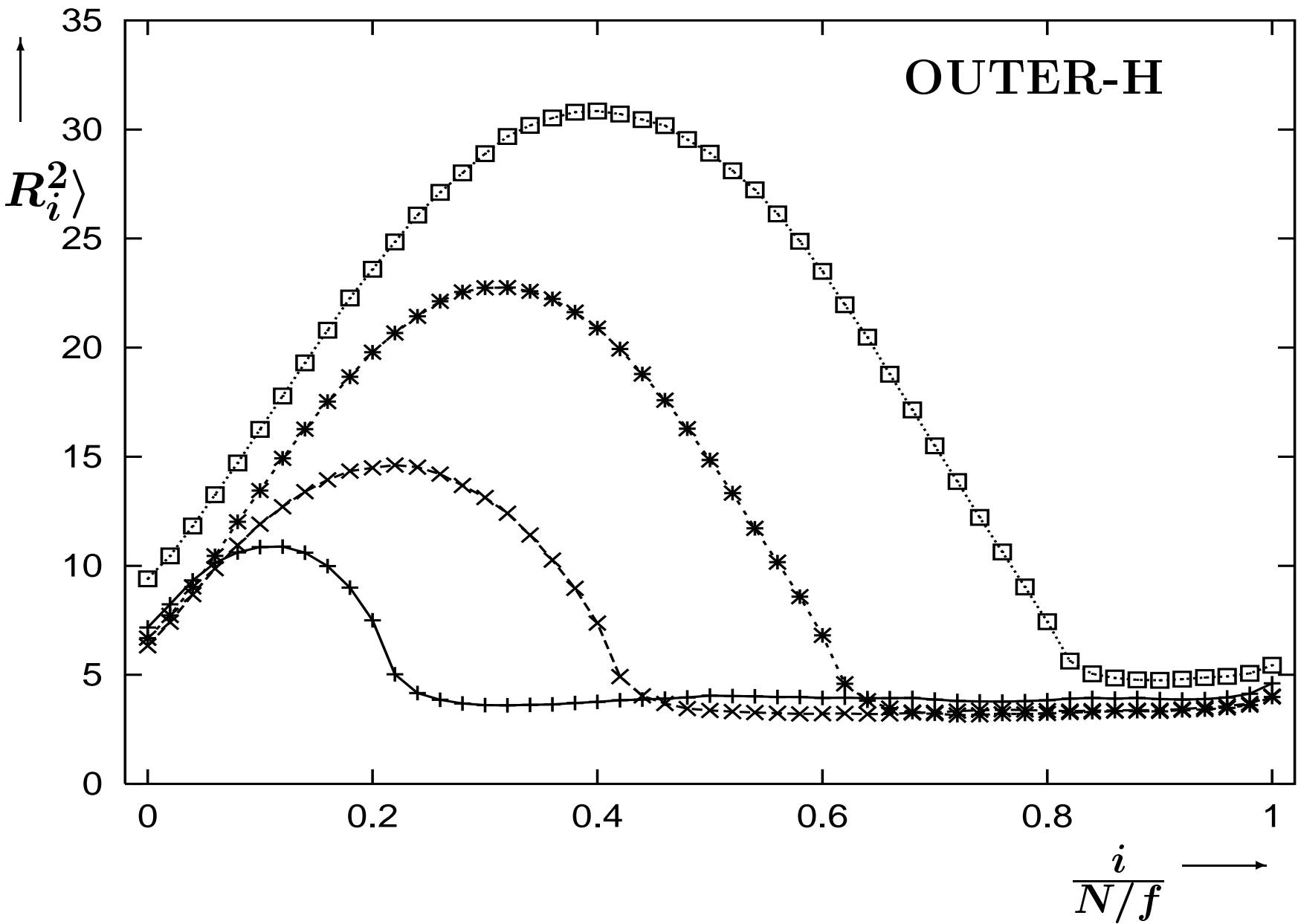
\begin{figure}
\caption{
\label{fig:as3}
The plots of $\left\langle R_i^2 \right \rangle$ vs $i/(N/f)$ drawn as
in Fig. \ref{fig:as2}, but for OUTER-H stars
in the stable state.
}
\end{figure}

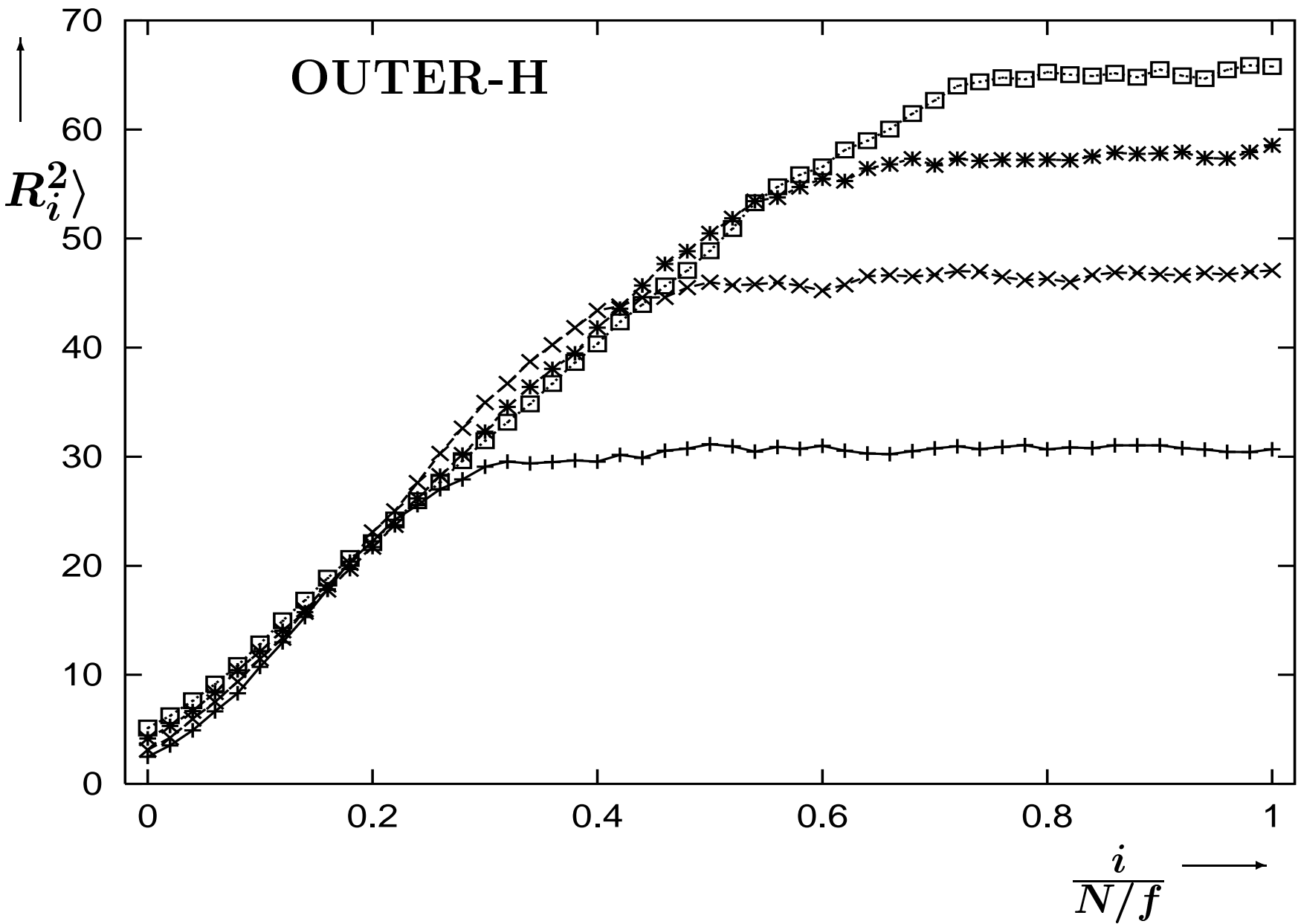
\begin{figure}
\caption{
\label{fig:as4}
The plots of $\left\langle R_i^2 \right \rangle$ vs $i/(N/f)$ drawn as
in Fig. \ref{fig:as2}, but for OUTER-H stars
in the metastable state.
}
\end{figure}

In Figs. \ref{fig:as2},\ref{fig:as3},\ref{fig:as4} we exhibit
the mean--squared distances of monomers from the centre of mass for
the INNER-H, OUTER-H and metastable OUTER-H states respectively.
Naturally, the shapes of these curves experience predictable
and monotonic shifts, reflecting the changes of the lengths of the 
H- and P-blocks. So, $\langle R^2_i\rangle$ for the INNER-H
structure remains nearly constant within the collapsed
H-core, starting to grow characteristically for the P-corona beads
with increasing chain index $i$. This function for the stable state of
OUTER-H stars has a familiar bell--shape for the P-units, forming loops
around the collapsed H-core, for which the function remains nearly
constant. And for the metastable state of the OUTER-H stars
we have a steady growth of $\langle R^2_i\rangle$ as long as
$i$ belongs to the P-block, which changes to a nearly constant
behaviour when $i$ reaches the H-block. Such types of behaviour are
fully consistent with the structures seen in snapshots for the symmetric
diblocks. It is also interesting to follow in more detail the
separation of the core monomer $i=0$ from the centre of mass.
For the INNER-H globules the core monomer remains very close
to the centre of mass similarly to the simple homopolymer star.
The same conclusion is also true for the metastable OUTER-H structures,
whereas for the stable OUTER-H state $\langle R^2_0\rangle$ is
quite large, especially for small $N_{H}/N_{Tot}$ ratios
due to long loops formed by the P-beads to all of which the
core monomer is connected.


\section{Conclusion}

In the present paper we have studied the behaviour 
of block copolymer stars in dilute solution by 
means of analytical Gaussian variational theory and 
Monte Carlo simulation in continuous space. 
We considered stars with each arm consisting of a diblock chain, 
one block being formed by hydrophilic (polar) monomers or P-beads, and
the other one by hydrophobic monomers or H-beads.
Moreover, we were interested in symmetric stars either having 
the H-block connected to the core bead and with the P-block 
being outside --- an INNER-H star, or having the opposite connectivity 
--- an OUTER-H star.  In both cases, we have found that for a sufficiently 
large degree  of amphiphilicity the stable state corresponds to
conformations with a dense globule 
formed by the H-beads, which is surrounded by the P-blocks 
stretching outwards.   However,
because of the intramolecular connectivity, the OUTER-H 
stars have a considerably smaller overall size than the INNER-H ones.
Indeed, the P-beads have to return back towards the globule of the H-beads
and they are all connected to the same core monomer.

Interestingly, for a given degree of amphiphilicity,
the size of the H-globule turns out to depend only on the 
number of H-beads, but it is  basically independent of the star 
topology (i.e. whether it is an INNER-H or an OUTER-H star), 
and only rather weakly dependent on the number of arms. 
The latter dependence is related to the 
repulsive interactions among the P-beads, whose number increases 
with the star functionality.

Another interesting result of our study was in finding that there is 
a metastable state for the OUTER-H stars. The conformation
corresponding to this state looks precisely as a swollen mini--star
with the H-beads clustering into small subglobules at the ends of the
outstretched P-blocks.
The repulsive interactions of the inner P-beads and an entropic barrier
prevent the coalescence of these small subglobules into a single globule, 
as it is the case in the stable state. Importantly,
this metastable state was also found by the Dynamic Monte 
Carlo simulation during kinetics after a sudden quench
from the swollen coil to the poor solvent regime.
This kinetic intermediary appears to be a rather
long--lived metastable state separated by a nucleation stage from
the final equilibrium state.  We have also seen that
this nucleation regime would transform
into a spinodal regime rendering such a state unstable during
kinetics after a much deeper quench (i.e. larger degree of amphiphilicity).

One may argue that the existence of such a metastable
state would have an obvious implication for the phase diagram 
of these stars. Indeed, the attractive interactions among the H-beads, 
which lead to the intramolecular formation of the small droplets at the 
ends of the arms, would also lead to star aggregation through intermolecular 
interactions and, eventually, to phase separation. 
Conversely, in the stable state, the inner globules of two molecules 
would not aggregate as easily because they are shielded by the outer P-beads 
exhibiting mutually repulsive interactions, which therefore act as colloidal 
stabilisers. 

We have also studied diblock stars with varying H:P composition ratio
and have found that the above general conclusions, such as e.g.
the existence of the OUTER-H metastable state, remain valid
for a very broad range of composition ratios. The family of plots
for the mean--squared distances of monomers from the centre of mass versus
the chain index has been systematically obtained, but it
uncovers only very predictable patterns. 

Given a certain technical distinction in the models adopted
by the analytical Gaussian theory and the Monte Carlo simulation,
as well as expected limitations of the former due to approximations involved,
we have found a very good agreement in the results from both
techniques.
This is a clear indication that the behaviour 
we are seeing is indeed universal, and it is not really affected
by the analytical approximations or implementation of
the simulation. 

Speaking of small discrepancies between the results, one
should mention the trapping of the core P-bead within
the H-globule in the stable state of OUTER-H stars as
seen by the analytical theory.
However, all simulation results indicate that this bead can 
actually escape from the globule, clearly leading to a more 
energetically favourable situation. 
The trapping is possibly enhanced by the 
small number of degrees of freedom for short 
lengths of the arms, but in any case is believed to be an artefact
of the theory.

Perhaps one of the most impressive agreements in the results
was obtained for the density profiles, as long as one used the reduced
variables $\hat{\rho}_a$ vs $\hat{r}$, with $a$ being $P$, $H$, 
or $Tot$. Even though the numbers of arms and their lengths differed
significantly in both considered cases, the plots in terms of the 
reduced dimensionless variables showed a robust consistency with each other.
This may seem unexpected given that the density can only be expressed
via a superposition of Gaussian--shaped functions within the analytical
theory.
One noticeable discrepancy with the simulation results was found, however, 
for $\hat{\rho}_P$ in the stable globular state of the INNER-H and 
OUTER-H stars within the range $\hat{r} \leq 1$.  In that region, the density
of the P-beads goes quickly to zero for $\hat{r} \leq 0.5$  in the results of
the simulation, showing a perfect $P$-beads  
expulsion from the space occupied by the dense H-subglobule.
This behaviour, however, manifests itself only in a relatively 
moderate density decrease in case of the Gaussian theory, which
on the other hand somewhat distorts the overall density shape due to 
the normalisation requirement.
Apart from this understandable difference, the overall shapes, peaks 
locations, heights and widths are in good agreement between the
theory and simulation.

The overall good agreement of the Gaussian variational theory
with the simulation in application to such complicated a system as a diblock
copolymer star, demonstrates the strength of the theory for predicting
the details of polymer conformations. The main idea of the Gaussian
variational theory is in determining the effective connectivity
matrix between all beads which would mimic the true intramolecular
interactions in the most optimal way, i.e. minimise the trial
free energy, ${\cal A}_{trial}={\cal A}_0+\langle H-H_0 \rangle_0$,
where $H_0$ and ${\cal A}_0$ are the trial Hamiltonian and free
energy associated with it.
The choice of variational variables differs somewhat in the two
versions of the theory: it involves products of bond vectors 
$\langle \bbox{l}_i \cdot \bbox{l}_j \rangle$ in the Milan version, while
the set of mean squared distances
$\langle (\bbox{X}_i - \bbox{X}_j)^2 \rangle$
between monomers in the Dublin version. However, as it is demonstrated
in the Appendix below, the resulting free energy expressions are
in fact the same, up to a trivial additive constant and some
rescaling of the parameters of the model.
In so far as the phase diagram and behaviour of
simple observables, such as e.g. radius of gyration, mean--squared
distances and density profiles, are concerned, the Gaussian self--consistent
approximation turns out to be quite accurate.
It is also clear that the variational theory should become increasingly
successful in dealing with polymers with complex intramolecular
architecture as compared to linear chain polymers precisely because
the conformation of the former polymers is more dictated by the
connectivity constraints. 

Furthermore, the current theory also permits extensions 
to the realm of kinetics
by writing time--dependent equations for instantaneous values of
the adopted averaged variables in terms of the instantaneous
free energy gradients (see. e.g. ref. \cite{e}).
Such future kinetic studies of copolymers with complex topology
present a considerable interest.
To illustrate this point, we may mention that folding of even
a homopolymer star proceeds considerably faster than of
the equivalent linear chain, and this is even more so for an INNER-H
star vs diblock linear chain. In certain cases of ``random'' copolymers
with a specially optimised complex topology one may achieve a 
much more efficient and faster folding than for the equivalent
linear chain, which would yet produce a globule
with an almost identical conformational structure since
the connectivity constraints are less relevant within the dense
globule. This may present certain new insights and 
opportunities for the broad areas of biopolymer folding and perhaps
even drug delivery.

Speaking of more traditional applications, we believe that the results
of this work may facilitate some further experimental research on
copolymer stars with a view of their applications such as as viscosity
modifiers and coating additives. 
On the one hand, the existence of distinct stable
compact and semi--extended conformational states with good colloidal
stabilisation properties seems of certain interest.
The presence of the semi--extended metastable state with ``sticky'' properties,
moreover, indicates that copolymer stars in solutions at
higher concentrations may exhibit a spectacular phase behaviour.
While this is interesting,
the studies of aggregation phenomena within the present very
detailed treatment are not particularly justified due to considerable
computational expenses. A somewhat more coarse--grained approach, both
analytically and in terms of the Monte Carlo model, may lead to
a faster progress in this direction. We expect that appropriate techniques
for such analysis could soon be developed and that the
corresponding complex phenomena will be thoroughly studied within
the next few years.
Finally, we would like to express a hope that the theoretical and
computational study of macromolecular solutions 
involving polymers with increasingly complex design,
as well as the collaboration between our two Groups, will continue in
the future.

\acknowledgments

The authors are thankful to Professor Giuseppe Allegra, Professor
Angelo Perico and Dr Guido Raos for interesting discussions.
One of us (E.T.) would like 
to thank Gavin McCullagh and Ronan Connolly for their work on 
the related GUI environment and gaining preliminary insights into the problem
during the course of their undergraduate projects.

This work was supported by grant IC/1999/01 from Enterprise Ireland,
which has helped to develop a fruitful collaboration between the 
Dublin and Milan Groups.

\appendix
\section*{Mathematical details and notes on the Gaussian method}

Comparison of the results from the theory and simulation using
different models of connectivity in this paper show that as far
as the universal properties of polymers are concerned one
could use either of the two models: freely--jointed or bead--and--spring
one. Since the connectivity term (\ref{cmc:hamil}) is the only one
which is affected by the molecular topology,
it would suffice to discuss the theory for a linear polymer chain,
in which one has to change the connectivity matrix appropriately.

The purpose of this Appendix is to give some mathematical details
on the analytical equations employed, but as an added bonus to
demonstrate the equivalence of the equilibrium free energy expressions
used independently by the two Groups.
Thus, let us establish an explicit relationship between the free energy of the
GSC method in ref. \cite{e} and that of refs. \cite{a,h}.

First of all, monomer coordinates can be written for a linear 
chain via the bond vectors as follows,
$\bbox{X}_i = \sum_{j=0}^i \bbox{l}_j$, if we  choose the
reference point as, $\bbox{X}_0=0$.
Then, the mean squared distances between monomers $i$ and $j$
are written as
\begin{equation}
D_{ij}\equiv \frac{1}{3} \langle (\bbox{X}_i-\bbox{X}_j)^2 
\rangle.
\end{equation}
Therefore, the entropic term (Eq. (7) in ref. \cite{h}) in units
of $k_B T=1$,
\begin{equation}
\label{APent}
{\cal A}_{el} = \frac{3}{2}\left( \mbox{Tr}\,\bbox{M} 
- \ln\mbox{Det}\,\bbox{M} \right),
\end{equation}
where we have discarded a constant independent of the variational
variables and the matrix $\bbox{M}$ defined as,
\begin{equation}
M_{ij} \equiv \frac{1}{l^{2}} \langle \bbox{l}_i \cdot \bbox{l}_j \rangle 
\end{equation}
can be rewritten as follows. 
Its first part becomes the harmonic springs ``energy'' 
(Eq. (10) in ref. \cite{e}),
\begin{equation}
\label{APspen}
{\cal E}_{sp}= \frac{3}{2\ell^2} \sum_{i\sim j} D_{ij}, \quad
\ell \equiv \frac{l}{\sqrt{3}},
\end{equation}
which in the latter form would also be valid for stars and any
other topology as it is precisely the first term in Eq. (\ref{cmc:hamil}) 
employed in Sec. \ref{subsec:cmc}.

Now, let us consider the determinant of the matrix $\bbox{M}$, where
we have scaled out the constant $l$,
\begin{equation}
\label{APdetM}
\left(
\begin{array}{cccc}
\langle\bbox{l}_1\cdot\bbox{l}_1 \rangle 
& \langle\bbox{l}_1\cdot\bbox{l}_2 \rangle 
&  \ldots & \langle \bbox{l}_1\cdot\bbox{l}_{N-1} \rangle\\
\langle\bbox{l}_2\cdot\bbox{l}_1 \rangle 
& \langle\bbox{l}_2\cdot\bbox{l}_2 \rangle
& \ldots & \langle\bbox{l}_2\cdot\bbox{l}_{N-1}\rangle\\
\ldots & \ldots & \ldots & \ldots \\
\langle\bbox{l}_{N-1}\cdot\bbox{l}_1 \rangle 
& \langle\bbox{l}_{N-1}\cdot\bbox{l}_2 
\rangle & \ldots & \langle\bbox{l}_{N-1}\cdot\bbox{l}_{N-1}\rangle
\end{array}
\right).
\end{equation}
Obviously,
the determinant of a matrix does not change if one adds a column 
or a row to any other columns or rows.
Thus, by adding the first row to the second, then the new second to
third and so on, we obtain,
\begin{equation}
\label{APdetM1}
\left(
\begin{array}{cccc}
\langle \bbox{X}_1\cdot\bbox{l}_1 \rangle 
&  \langle\bbox{X}_1\cdot\bbox{l}_2 \rangle 
&  \ldots & \langle \bbox{X}_1\cdot\bbox{l}_{N-1} \rangle\\
\langle \bbox{X}_2\cdot\bbox{l}_1 \rangle 
& \langle\bbox{X}_2\cdot\bbox{l}_2 \rangle 
& \ldots & \langle \bbox{X}_2\cdot\bbox{l}_{N-1} \rangle\\
\ldots & \ldots & \ldots & \ldots \\
\langle\bbox{X}_{N-1}\cdot\bbox{l}_1 \rangle 
& \langle \bbox{X}_{N-1}\cdot\bbox{l}_2 \rangle
 & \ldots & \langle \bbox{X}_{N-1}\cdot\bbox{l}_{N-1} \rangle
\end{array}
\right).
\end{equation}
By repeating the same procedure for the columns, likewise we conclude
that 
\begin{equation}
\mbox{Det}\,\bbox{M}=\mbox{Det}\,\bbox{{\cal F}} , \qquad
{\cal F}_{ij}\equiv \frac{1}{l^{2}} \langle \bbox{X}_i\cdot\bbox{X}_j \rangle, 
\end{equation}
which in turn agrees with Eqs. (B3) and (B5) in ref. \cite{e} upon
fixing the reference point at $\bbox{X}_0=0$.

Finally, the virial expansion of the excluded volume energy
(Eq. (10) in ref. \cite{e}),
\begin{equation}
\label{APee}
{\cal E}_{exc}=
\sum_{J=2}^{\infty}\sum_{\{i\}}B^{(J)}_{\{i\}}({\rm Det}
\bbox{\Delta}^{(J-1)})^{-3/2}, \qquad
\Delta^{(J-1)}_{rs} \equiv \frac{1}{3}\left\langle
\left(\bbox{X}_{i_1}-\bbox{X}_{i_{r+1}}\right)\cdot
\left(\bbox{X}_{i_1}-\bbox{X}_{i_{s+1}} \right)\right\rangle,
\end{equation}
expands into the following explicit formulae for the determinants appearing
in the two-- and three--body interaction terms,
\begin{eqnarray}
{\rm Det} \bbox{\Delta}^{(1)}_{ii'}    & = & D_{ii'} \label{gsc:det2b} \\
{\rm Det} \bbox{\Delta}^{(2)}_{ii'i''} & = &
  \begin{array}{|cc|}
  D_{ii'} & \frac{1}{2}\left( D_{ii'} + D_{ii''} - D_{i'i''} \right) \\
\frac{1}{2}\left( D_{ii'} + D_{ii''} - D_{i'i''} \right) & D_{ii''}
  \end{array} = \nonumber \\
 & & = \frac{1}{2}\left( D_{ii'} D_{ii''} + D_{ii'} D_{i'i''} +
    D_{ii''} D_{i'i''} \right) - \nonumber \\
 & & - \frac{1}{4} \left( D_{ii'}^2 + D_{ii''}^2 + D_{i'i''}^2 \right).
\label{gsc:det3b}
\end{eqnarray}
These are precisely the expressions for the
two--body and three--body interaction energy terms as in
Eqs. (50,51) of ref. \cite{h}.
The positiveness of the first determinant requires all
mean squared distances to be positive, $D_{ii'} > 0$,
which is true as this is the average of a quantity squared.
The positiveness of the three--body determinant in
Eq.~(\ref{gsc:det3b}), on the other hand,
is insured by the triangle inequality,
\begin{equation}
D_{ii''}^{1/2} < D_{ii'}^{1/2} + D_{i'i''}^{1/2}.
\end{equation}
It is interesting to note that
the quantity $({\rm Det} \bbox{\Delta}^{(J-1)}_{i_1,\,\ldots,\, i_J})^{3/2}$
characterises the volume of the phase space occupied by $J$ particles
if the mean squared distances between all pairs are specified. Moreover,
the logarithm of this characteristic volume of all
$N$ particles is equal to the variational
entropy of the system up to a trivial constant,
\begin{equation}
{\rm Det} \bbox{M} = \frac{1}{N^2} {\rm Det} \bbox{\Delta}^{(N-1)}.
\end{equation}
This completes establishing the equivalence of the variational
free energy expressions between the methods --- one which is based
on bond--vectors \cite{a,b,g,h} and another which is based on the monomer
coordinates and mean squared distances between monomers
\cite{TorusNew,c,d,e,f}.



\newpage

\begin{figure}
\caption{ 
\label{fig:fc}
The mean--squared distances of the beads from the common centre of mass, 
$\left\langle R_i^2 \right \rangle$, (in units of the bond length $l$) 
of the 3-arm INNER-H star at $\sigma=1.0$ (part a, on the left) 
and OUTER-H star (part b, on the right). The lower solid curve corresponds to 
the stable state and the upper dashed curve to the metastable state,
plotted as a function of the bead position along an arm. Here
$i=0$ is the branch point, $i=N/f$ is the free end.  
Only the beads of one arm are shown due to the molecular symmetry 
obtained in all states.  The black circles indicate the H-beads and 
the white circles the P-beads. The vertical dotted line 
at $\frac{i}{N/f} = 0.5$ indicates the last bead of the inner block.  
}
\end{figure}

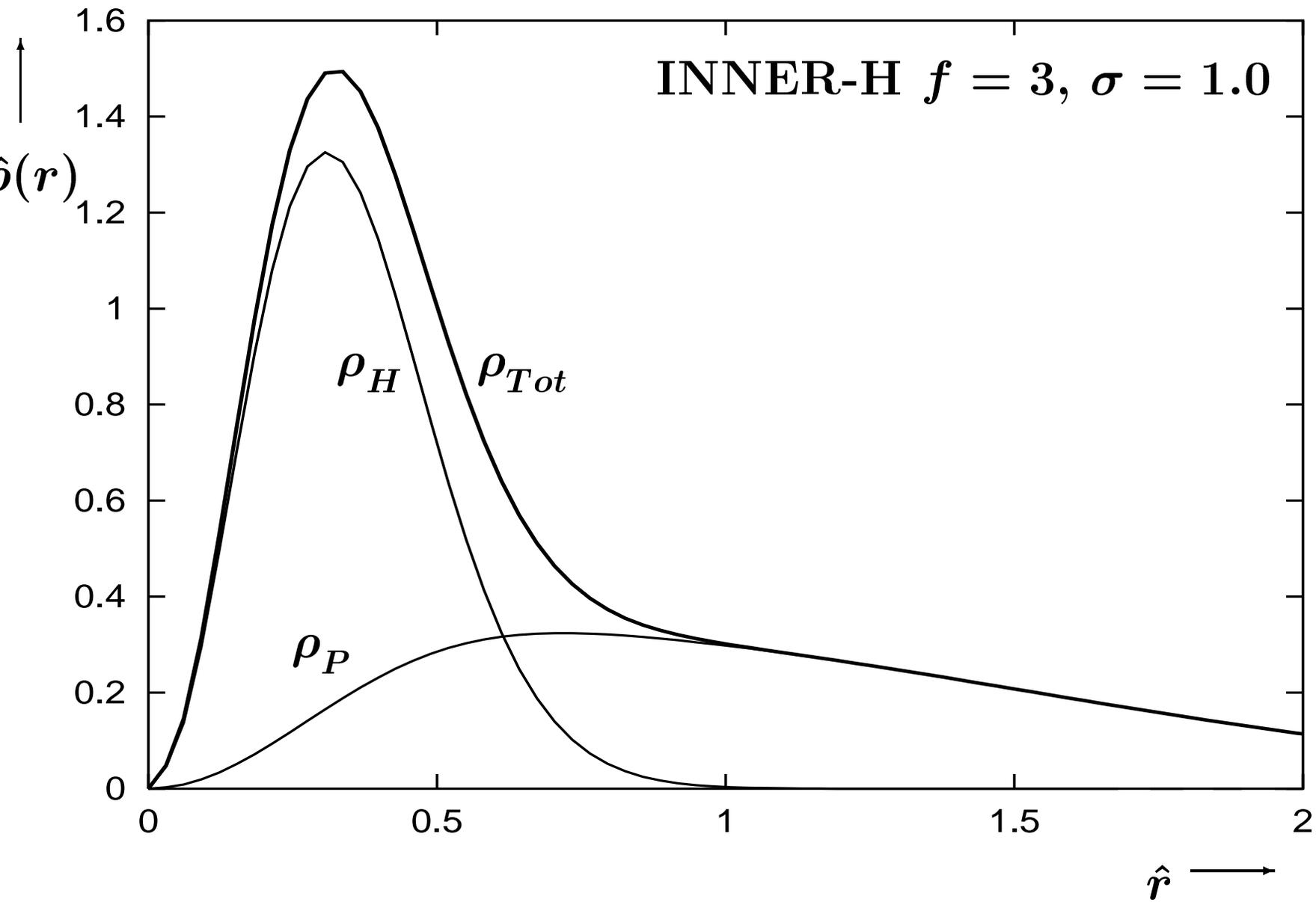
\begin{figure}
\caption{ 
\label{fig:fd}
The dimensionless reduced density $\hat{\rho}(r)$ profiles 
vs the dimensionless distance $\hat{r}$ (see Eq. (\ref{rhodef})) 
for the same INNER-H star as in Fig. \ref{fig:fc}a. 
These densities are drawn for the H-beads, P-beads and all beads.
The normalisation condition for each of these functions are simply, 
$\int_{0}^{\infty}d\hat{r}\,\hat{\rho}_a(r)=\frac{N_a}{N_{Tot}}$.
}
\end{figure}

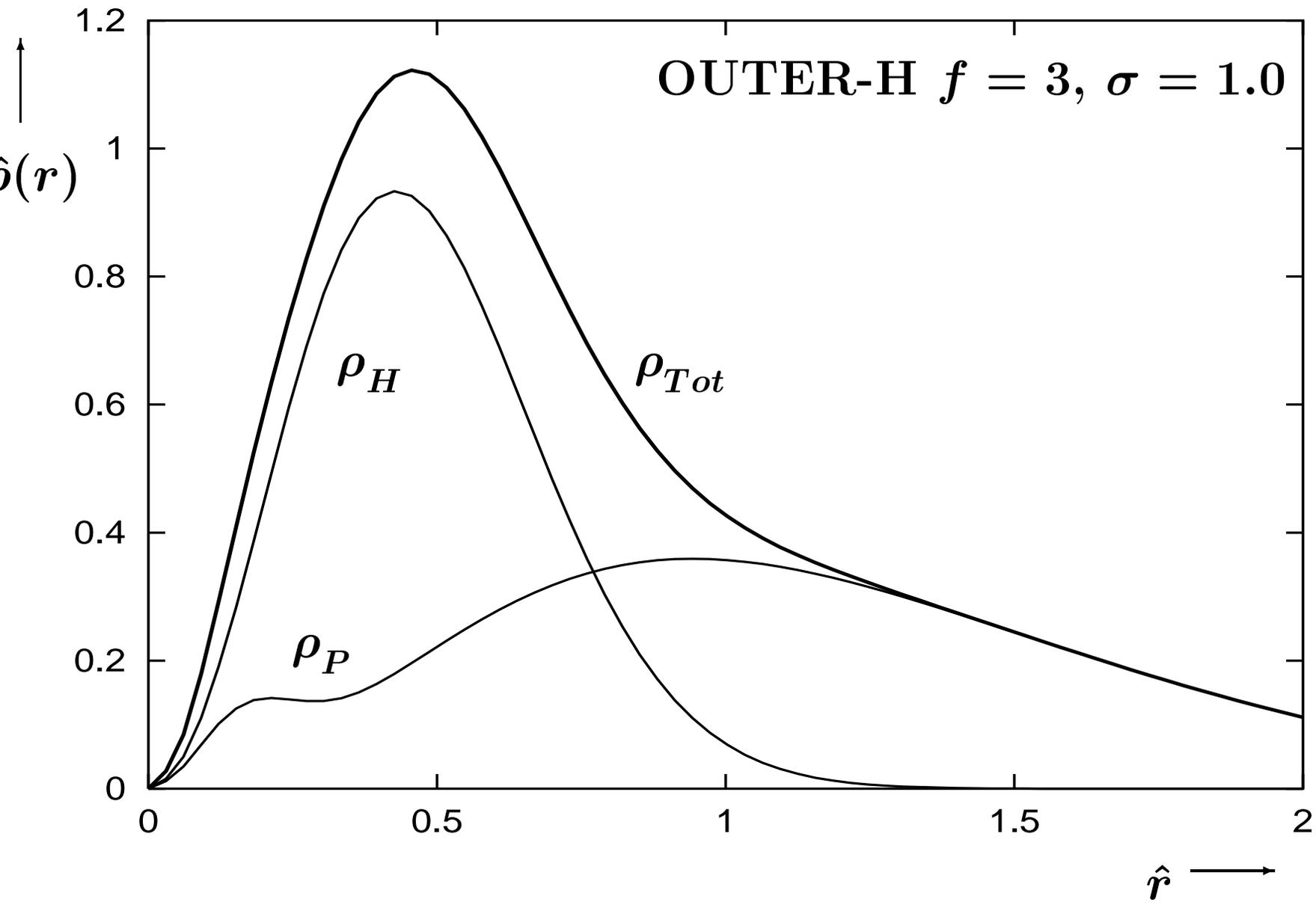
\begin{figure}
\caption{ 
\label{fig:fe}
The dimensionless density profiles 
vs the dimensionless distance for the 3-arm OUTER-H star in 
the stable state at $\sigma$ = 1.0 
drawn as in Fig. \ref{fig:fd}.  
}
\end{figure}

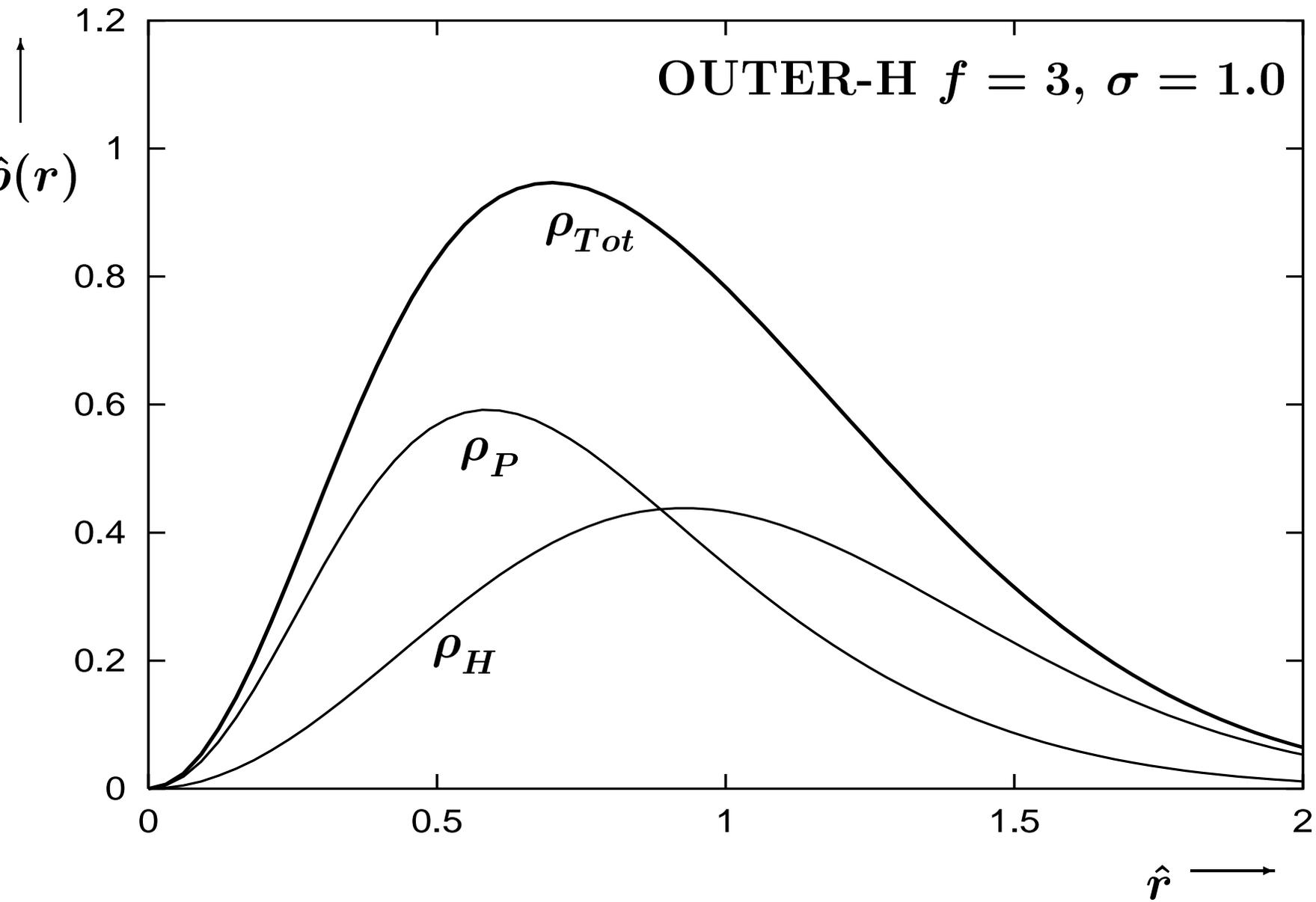
\begin{figure}
\caption{ 
\label{fig:ff}
The dimensionless density profiles 
vs the dimensionless distance for the 3-arm OUTER-H star 
in the metastable state at $\sigma=1.0$ drawn as in Fig. \ref{fig:fd}.
}
\end{figure}

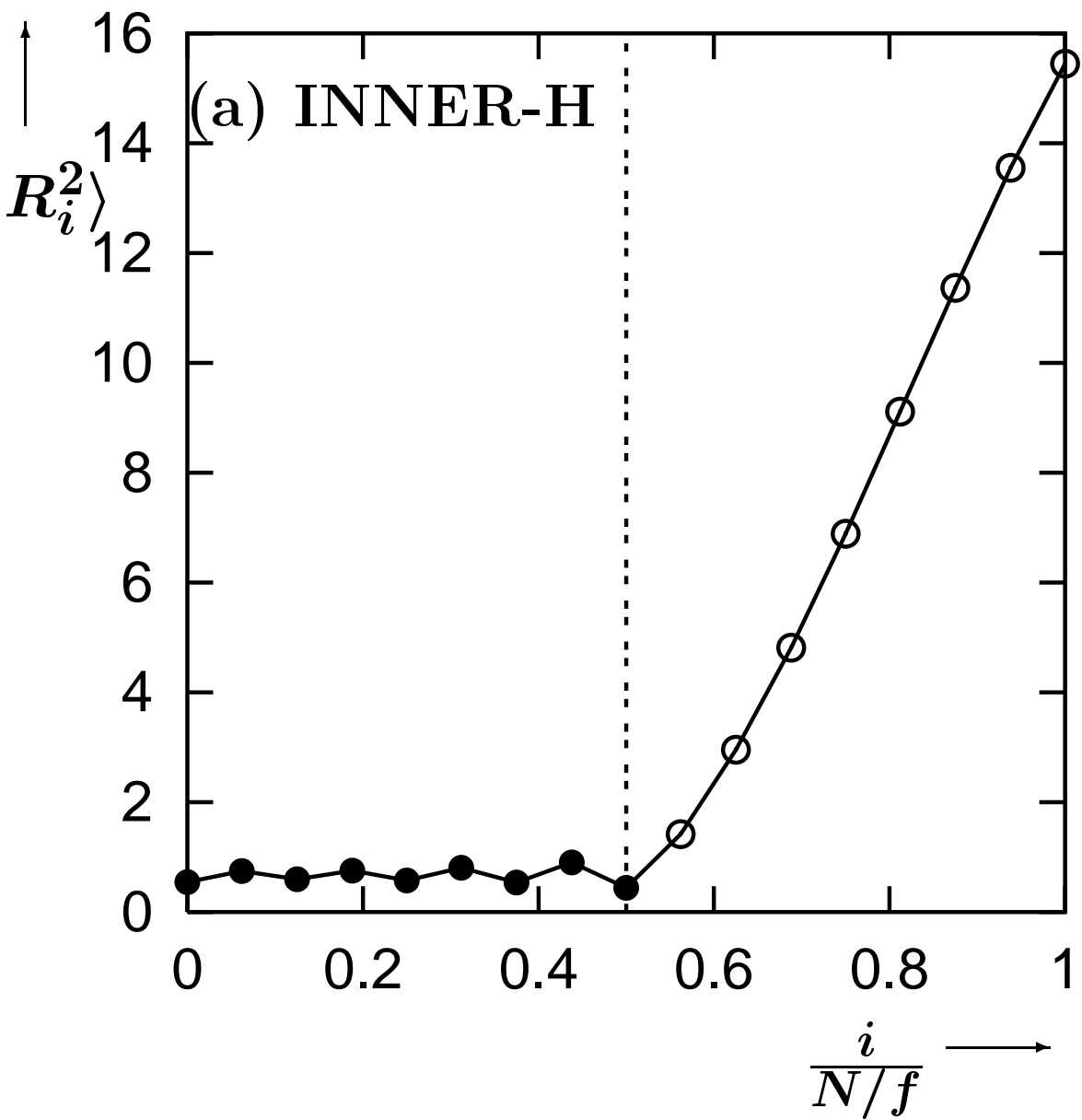
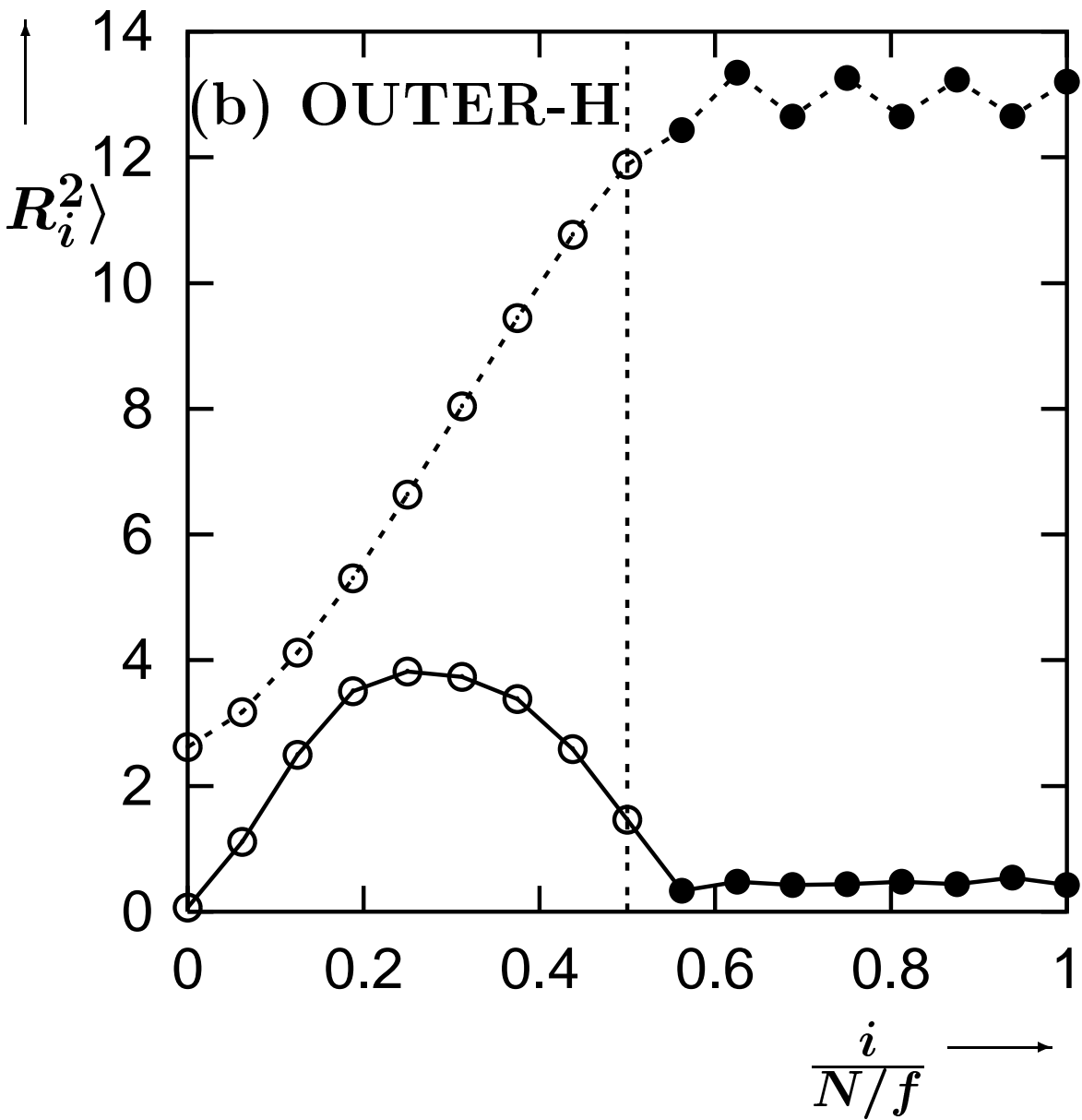
\end{document}